\documentclass[11pt,a4paper]{article}

\pdfoutput=1
\usepackage{jstyle,feynmf}

\usepackage{mathpazo}
\usepackage{environ}
\usepackage{mathrsfs}
\usepackage{array,arydshln}

\author[a,b]{Euihun Joung,}
\author[c]{\quad Simon Nakach,}
\author[c]{\quad Arkady A. Tseytlin}

\affiliation[a]{School of Physics and Astronomy, %  and Center for Theoretical Physics
Seoul National University, Seoul 151-747,  Korea}
\affiliation[b]{Gauge, Gravity \& Strings, \ Center for Theoretical Physics of the Universe,\\
Institute for Basic Sciences, Daejeon 34047, Korea}
\affiliation[c]{Theoretical physics group, Blackett Laboratory, Imperial College London,   SW7 2AZ, U.K. }

\emailAdd{euihun.joung@snu.ac.kr}
\emailAdd{nakach@ic.ac.uk}
\emailAdd{tseytlin@ic.ac.uk}

\def \fo {{\textstyle { 1 \ov 4}}}

\def \lab {\label}
\newcommand \foot [1] {\footnote{#1\vspace{2pt}}}
\newcommand \rf [1] {(\ref{#1})}

\def \ha {{1 \over 2}}
\def \td {\tilde}
\def \ci{\cite}

\def \del{ \partial}
\def\ov{\over}
\def\no{\nonumber} 

\def \ci {\cite}
\def \p {\phi}
\def \m {\mu}\def \n {\nu} 
\def \te {\textstyle} 
\newcommand{\me}{\mathfrak{e}}
\newcommand{\mh}{\mathfrak{h}}

\usepackage[nodayofweek]{date time}

\begin{document}

\date{\currenttime}
%%%%%%%%%%%%%%%%%%%%%%%%%%%%%%%%%

\title{\centering
\LARGE{ Scalar scattering\\  via   conformal higher spin   exchange}}

\abstract{Theories  containing infinite number of higher spin fields 
require a particular definition of summation over spins  consistent with their underlying symmetries. 
%Motivated by the  question of understanding this summation prescription 
%the question of choice of  definition of summation  over spins   in 
% theories  containing infinite towers of higher spin
%fields, 
  We  consider a   model of   massless scalars   interacting  (via  bilinear conserved currents) 
with   conformal higher  spin  fields in flat space.
  We   compute the tree-level four-scalar   scattering amplitude 
  using a natural prescription  for summation over  an infinite  set of conformal higher spin exchanges 
and  find that it vanishes 
%v3
 (modulo delta-function terms having  support on measure-zero domain in phase space).
 Independently, we   show  that this  vanishing  of the   scalar scattering amplitude is, in fact, 
implied  by the global conformal   higher spin symmetry of this  model. 
%This implies that the regularization 
%of summation over spins  we used is consistent with the underlying symmetry. 
We also discuss   one-loop corrections to the  four-scalar  scattering 
amplitude. %   assuming that  interacting conformal higher spin theory is defined as an induced theory.
%(with   conformal higher spins appearing on two of four internal lines)
%and show that its 0-momentum limit vanishes  in the same  summation over spins  regularization, 
%implying that one does not need to introduce  local 4-scalar counterterm in this model. 
}

\today
\date {}
\begin{flushright}\small{Imperial-TP-AT-2015-{04}}\end{flushright}

\maketitle

\def \iffa {\iffalse}

 \iffa 
 %%% Submission data  %%%%%
%%%%%%%%%%%%%%%%%%%%%%%%%%%%%%%%%%%%%%
Authors:  E. Joung, S. Nakach, A.A. Tseytlin
%%%%%%%%%%%%%%%%
Report No.: Imperial-TP-AT-2015-04
Title:     Scalar scattering via  conformal higher spin exchange
Comments:    pages. 
Abstract: 
Theories  containing infinite number of higher spin fields 
require a particular definition of summation over spins  consistent with their underlying symmetries. 
  We  consider a   model of   massless scalars   interacting  (via  bilinear conserved currents) 
with   conformal higher  spin  fields in flat space.
  We   compute the tree-level four-scalar   scattering amplitude 
  using a natural prescription  for  summation over  an infinite  set of conformal higher spin exchanges 
and  find that it vanishes
%v3 
(modulo delta-function terms having  support on measure-zero domain in phase space).
 Independently, we   show  that this  vanishing  of the   scalar scattering amplitude is, in fact, 
implied  by the global conformal   higher spin symmetry of this  model. 
%This implies that the regularisation 
%of summation over spins  we used is consistent with the underlying symmetry. 
We also discuss   one-loop corrections to the  four-scalar  scattering 
amplitude.

 \fi 

\def \ko {\kappa} 

\def \N {{\rm N}}
\def \OO {{\cal J}}
\def \O {{\cal O}} 
\def \rP {{\rm P}}
\def \edo {  \end{document} }
\def \chh {  \phi} 
\def \o {w}

%%%%%%%%%%%%%%%%%%%%%%%%%%%%%%
\section{Introduction}
\label{intro}
%%%%%%%%%%%%%%%%%%%%%%%%%%%%%%%

Higher spin theories containing infinite number of  particles  pose a challenge  of how to define  them 
at the quantum level in a way consistent with their large amount of symmetry. One particular 
issue is how to treat sums over infinite number of spins. 
This   question was recently   addressed  on   examples 
of simplest  higher spin partition functions  in \ci{Beccaria:2015vaa} following 
 \ci{%Gupta:2012he,
 Giombi:2013yva,Giombi:2013fka,Tseytlin:2013jya,Giombi:2014iua, Giombi:2014yra,
 Beccaria:2014jxa,Beccaria:2014xda}.
 
 Our aim will be to  study  this issue in the   context  of  S-matrix   of   scalars  interacting 
 via  exchange of an infinite set  of  higher spin  fields. 
  This   is an analog of  the  Veneziano amplitude  in string theory 
 where   the  infinite tower of exchanged fields  are  massive. 
 This set-up     was originally discussed in  \ci{Bekaert:2009ud}  where a tree-level scalar   scattering amplitude  with  standard massless 
 higher spin particles  exchange  was  considered. 
   Since an  interacting theory of massless  higher spin particles ought  to be  not well-defined  in  flat space (cf. \ci{Weinberg:1964ew,Aragone:1979hx,Fradkin:1987ks})  the computation of    \ci{Bekaert:2009ud}  
  is, however,  hard to  embed into  a consistent theory.

 Here instead   we  shall  consider  a  model where  the  scalars  interact through exchange of a tower of 
 {\it  conformal higher spin}  fields. 
 Conformal Higher  Spin (CHS)  theories  are generalisations  of $d=4$ Maxwell  ($s=1$) 
and Weyl ($s=2$) theories  that describe  pure spin $s$ states  off shell, i.e.  have  maximal  gauge  
symmetry  consistent with  locality at the expense of having higher-derivative kinetic terms \ci{Fradkin:1985am} 
(see also  \ci{Fradkin:1989md,Tseytlin:2002gz,Segal:2002gd,Bekaert:2010ky,Beccaria:2014jxa}). 
In contrast to the   two-derivative massless  higher spin theory, 
the CHS theory  (that can be defined at the full non-linear level   as 
 the UV singular  local part of the  induced action of free scalars  with 
  higher spin background  fields  coupled to 
 all conserved spin $s$ scalar currents \ci{Tseytlin:2002gz,Segal:2002gd,Bekaert:2010ky}) 
   may be viewed   as a  formally consistent 
(but a priori non-unitary)  
 interacting gauge theory   when expanded  near  flat space. 

 To  introduce a  particular  model which we shall study  in this paper,  let us first recall the basics  of  vectorial AdS/CFT  duality 
 (see, e.g., \ci{Klebanov:2002ja,Giombi:2014yra,Beccaria:2014xda}). 
 Consider a free  CFT$_d$  of $N$  complex scalar fields  
 \be\lab{1a}
 S=\int d^d x\  \vec\chi^*\!\cdot\del^2\vec\chi\,,
 \ee
with  primary conformal   operators  being  on-shell-conserved  traceless 
currents $J_{\m_1...\m_s}$ of dimension $\Delta= d-2 + s $. The latter 
 are  bilinear  $U(N)$ singlets  (see \ci{Craigie:1983fb})
 \be\lab{1b} 
 	J_s(\vec\chi)  = \vec\chi^*\!\cdot\OO_s\,\vec\chi\sim 
 \vec\chi^*\!\cdot\del^s\,\vec\chi\,, \qquad \qquad  s=0,1,2,\ldots \ , 
\ee
 where 
$\OO_s$ is an appropriate differential operator.
Introducing source   fields   $h_s(x)$   for   all  $J_s$  and integrating out 
$\vec\chi$\,, one gets a generating functional 
 for connected correlators of all currents
\be\lab{13}
	\G[h]= N \,\log \det ( -\del^2   + \sum_s h_s\,\OO_s)\,.
\ee
The $d$-dimensional  fields $h_s$  may be  viewed  as gauge fields for  
the symmetries of the free classical  scalar theory 
with  linearised  differential 
and  algebraic (``trace shifting") symmetries  
generalising  the reparametrization and  Weyl symmetry of the Weyl gravity. They  
   can   thus  be   identified   with  the CHS  fields.\foot{Demanding invariance under non-linear  symmetries   for a particular subset of fields 
   may  
 require introducing  extra terms non-linear in $h_s$ (like in scalar electrodynamics or in covariant coupling 
 to a curved metric).   However, being   local (involving powers of $h_s$ fields at the same point), they 
 would   not  change the values of  the 
 CFT  correlators of primary operators $J_s$  at separated points.}
 
 The same  functional $\G[h]$ \rf{13} should  follow  from the Vasiliev's  massless higher spin theory \cite{Vasiliev:1990cm,Vasiliev:1990en,Vasiliev:2004qz} in AdS$_{d+1}$ 
upon   integrating over the AdS$_{d+1}$   Fronsdal fields $\Phi_s$ with Dirichlet boundary conditions 
(\mt{\Phi_s\big|_{\del \rm AdS} = h_s}).  
The number of scalars $N$  then plays  the role of the  inverse coupling  of the  higher spin 
theory in   AdS$_{d+1}$   (appearing in front of its  classical  action).
 All quantum (order $N^0, N^{-1}, \ldots$) corrections to the generating functional  computed from the 
   Vasiliev's  theory   should  then vanish  to match the   boundary theory 
   result.\foot{More precisely, 
   what should vanish
    are corrections to derivatives of the generating functional  at separated points.} 

%AT
The  quadratic term  in $h_s$ term  of  $\G[h]$ in \rf{13}   is  
\be \lab{13a}
	\G_{2}[h]=N\,\sum_s \int h_s  K_s  h_s\,,  \\
\ee 	with 
\be \lab{14a}
	K_s 
\sim   N^{-1} \la J_s(x)   J_s(x') \ra \sim  \rP_s\,|x-x'|^{ 4-2d - 2s} 
\sim \rP_s\,\del^{2s + d-4}\,\delta^{(d)} (x-x')\, \log \Lambda +
\ldots
\ee
where $\rP_s$ is the transverse traceless projector and  $\Lambda$ is  a  UV cutoff. From now on, we assume $d$ is even. 
 Thus 
 the  UV singular part of $\G_2$  is proportional to  the 
 collection of  CHS kinetic terms 
$\int d^d x \ h_s\, \rP_s \, \del^{2s + d-4}\, h_s$\,. 

% why induced cubic hhh interactions are local?

Suppose   now   we start with $N+1$  scalar fields, $\vec\chi$ and $\p$, 
couple them  to  the CHS  fields $h_s$  via  the  currents $J_s(\vec\chi)+J_{s}(\p)$ 
and   integrate out only $N$ scalars  $\vec\chi$. 
 The resulting  effective  theory   will  contain  the remaining scalar  $\p$
    coupled  to the CHS   fields $h_s$  described  by the induced action, i.e.
  \be 
  S[ \p,h] =    \int d^d x \Big[  \p^* \del^2  \p  +   \sum_s  h_s\, J_s(\p) \Big]   + \G[h]\ , 
     \lab{1}
     \ee 
     where  $ \G[h]= N \sum_s  [\int  h_s K_s h_s +  \O (h^3) ] $. 
     The UV   singular local part of $ \G[h]$    may  be identified with a  non-linear   CHS action
     \ci{Tseytlin:2002gz,Segal:2002gd,Bekaert:2010ky}. 
     One may then 
      compute  the  S-matrix   for $\p$   due to the exchange of  the tower of  all CHS fields $h_s$.
   %  and try to understand the role of regularisation of  the sums over spins. 
     Assuming $N$ (or the  inverse CHS theory coupling) 
       is large   we may treat   self-interactions of $h_s$   in perturbation theory. 
     
      While   a non-trivial  S-matrix for $\p$ is  not a natural  observable  in  the boundary CFT$_d$ 
     (which is a free  theory from the start) this set-up is in a 
       sense a higher spin theory analog  of   the  computation of 
     the   4d gluon S-matrix  from the AdS$_5$   point of view  \ci{Alday:2007hr} where  
     one  first 
              ``integrates out"  $SU(N)$  gauge vectors  to ``{build}"   the bulk  geometry, and then 
              considers the   scattering of  extra  gluons on a   probe 3-brane. 

     In general,  one   may  study    the case when the CHS part $\G[h]$  of  the model \rf{1} 
     is given  by either   the full   non-local induced action (i.e. with  kinetic term 
     $\rP_{s}\,\del^{2s+ d-4} \log (\del^2/\Lambda^2)$)
      or simply   its  local  UV singular part 
      $\rP_{s}\,\del^{2s+ d-4}\,\log \Lambda$.
       The latter    choice  is %may be advantageous  
      preferable    when trying 
      to  include  also self-interactions  of $h_s$:   
       the finite part of  the full induced action is a priori anomalous,   breaking the  classical algebraic
       symmetries of  the CHS fields.\foot{The anomalous  part of the effective action does not, however, 
         contribute to the  
       correlation functions of   conformal current operators at separated points (the anomaly  expressions 
       contain at least two fields at the same  point).
       For example, 
       a scalar  $\phi$   coupled to the  background  metric $g_{\m\n}=\eta_{\m\n} + h_{\m\n}$  in a  reparametrisation and Weyl-covariant way (i.e. with 
       \mt{ { (d-2)\ov 4 (d-1) }\,R\,\phi^*\phi}  term included)   has Weyl-anomalous (starting with  cubic $(h_2)^3$ order) effective action  
       but its  UV divergent  part  $\sim$ (Weyl tensor)$^2$   is Weyl-invariant to all orders.
       }
       At the same  time, the   local $\log \L$ part of $\Gamma[h]$ is  
          invariant 
             under the %full non-linear 
            symmetries  of the CHS 
            theory   \ci{Segal:2002gd,Bekaert:2010ky}.
            %\foot{In general, the symmetry may seem  to 
          %  require non-linear in $h_s$  terms  to  be  included in the coupling of $\phi$ 
          %  and $h_s$ but they can be absorbed into a local field redefinition of $h_s$
          %  that will not change  correlators at separated points.
        %      } 
              
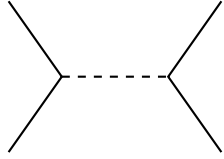
\begin{figure}[h]
\centering
\begin{tikzpicture}
\draw [thick] (0,0) -- (0.7,1) -- (0,2);
\draw [thick,dashed] (0.7,1) -- (2.1,1);
\draw [thick] (2.8,0) -- (2.1,1) -- (2.8,2);
\end{tikzpicture}
\caption{Tree-level diagram}
\label{fig1}
\end{figure}

              %NEW 
              In what follows we shall study 
     % Thus  the  investigation  of
       the  model   \rf{1} 
       viewed  as  a {\it  local}  CHS   theory  interacting  %(in fully covariant non-linear way) 
      with a free conformal scalar matter, i.e. assume that only the  local part of $\Gamma[h]$ 
      defining the CHS action $S[h]$  is   kept with coefficient $\k \sim N$  as the (inverse)  coupling  
      constant. 
      Starting with \rf{1} and rescaling $h_s $ 
      as   $h_{s}\to \sqrt{N}\,h_{s}$\,,   we get 
      \be \lab{2} S[\p,h]= 
       \int d^d x \Big[  \p^*\,\del^2\,\p  +  
         \sum_s    h_s\, \rP_s \, \del^{2s + d-4} \,h_s + 
     {\te  \frac1{\sqrt \k}}
       \Big(\sum_s  h_s\, \p^* \OO_s\,\p   +   
     	h^3 \Big)+  \O({\te {1\ov \k}}\,h^4)
         \Big] . \ee 
Thus at  the leading \mt{1\ov \k}   order  we get the four-scalar 
 tree level diagram (Fig.\ref{fig1}) with two \mt{1\ov \sqrt{\k}} vertices. 
Here the solid line  (---) stands for the scalar $\p$ propagator and the dashed line
(-\,-\,-) for  all CHS propagators. 
We shall explicitly compute the corresponding amplitude below. 

\begin{figure}[h]
\centering
 \begin{tikzpicture}
\draw [thick] (0,0) -- (0.7,0.5) -- (0.7,1.3) -- (0,1.8);
\draw [thick,dashed] (0.7,0.5) -- (2.1,0.5);
\draw [thick,dashed] (0.7,1.3) -- (2.1,1.3);
\draw [thick] (2.8,0) -- (2.1,0.5) -- (2.1,1.3) -- (2.8,1.8);
\end{tikzpicture}
\caption{Box diagram}
\label{fig2}
\end{figure}
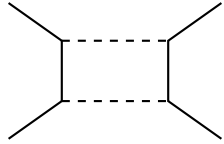

In addition, we shall  also  discuss the one-loop corrections to  4-scalar   scattering. 
  An example   of  such one-loop 
  order $1\ov \k^{2}$    diagram is the 1-PI one (Fig.\ref{fig2}) 
    with  four  \mt{1\ov \sqrt{\k}} vertices.
    %\foot{There  could  be 
   %   additional one-loop order $1\ov \k^2$ 1-PI  diagrams if one were to include non-linear 
%$h^2\,\chh^2$  coupling terms in \rf{1}.}
 The one-loop four-scalar amplitude of order $1\ov \k^2$ 
 receives   also contributions  from non-1-PI diagrams  which are 
  the tree-diagrams in Fig.\ref{fig1}  where 
 the  scalar legs, the CHS propagators and the vertices
 get the $1\ov \k$ corrections due to
 the scalar  self-energy diagram,   %(Fig.\ref{fig3}),
 the CHS  self-energy diagram,  %(Fig.\ref{fig4})
 and  the charge-renormalization diagram. % (Fig.\ref{fig5}), respectively.
 %%%%%%%%%%%%%%%%%%%
 
 \iffa 
\begin{figure}[h]
\centering
\begin{tikzpicture}
\draw [thick] (0,0) -- (4,0);
\draw [thick,dashed] (2.77,0) arc [radius=0.77, start angle=0, end angle= 180];
\end{tikzpicture}
\caption{Scalar self-energy}
\label{fig3}
\end{figure} 

\begin{figure}[h]
\centering
\begin{tikzpicture}
\draw [thick,dashed] (0.5,1) -- (1.4,1);
\draw [thick,dashed] (2.6,1) -- (3.5,1);
\draw [thick] (2,1) circle [radius=0.6];
\end{tikzpicture}
\qquad
\begin{tikzpicture}
\draw [thick,dashed] (0.5,1) -- (1.4,1);
\draw [thick,dashed] (2.6,1) -- (3.5,1);
\draw [thick,dashed] (2,1) circle [radius=0.6];
\end{tikzpicture}
\qquad
\begin{tikzpicture}
\draw [thick,dashed] (0.5,0) -- (3.5,0);
\draw [thick,dashed] (2,0.6) circle [radius=0.6];
\end{tikzpicture}
\caption{CHS self-energy}
\label{fig4}
\end{figure}

\begin{figure}[h]
\centering
\begin{tikzpicture}
\draw [thick] (0,0) -- (1.2,0.8) -- (0,1.6);
\draw [thick,dashed] (1.2,0.8) -- (2.2,0.8);
\draw [thick,dashed] (0.5,0.35) -- (0.5,1.25);
\end{tikzpicture}
\qquad\qquad 
\begin{tikzpicture}
\draw [thick] (0,0) -- (0.5,0.35) -- (0.5,1.25) -- (0,1.6);
\draw [thick,dashed] (0.5,0.35) -- (1.2,0.8);
\draw [thick,dashed] (0.5,1.25) -- (1.2,0.8) -- (2.2,0.8);
\end{tikzpicture}
\caption{Charge renormalisation}
\label{fig5}
\end{figure}

\fi

%Below  we shall mostly concentrate on the basic tree-level  4-scalar diagram in Fig.1 
%and  limit the  investigation of  loop corrections to the   study of the
 %local $(\chh^* \chh)^2$ term in the   one-loop  1-PI diagram in Fig. 2. 

We  will  start  in section \ref{sec2} with a   description of the model of a free  scalar field 
coupled to a tower of CHS fields. 
 In section \ref{sec3} we will compute the tree level amplitude  corresponding to 
 Fig.\ref{fig1}  using a particular regularisation prescription 
  for the sum over all spins. The resulting amplitude  will have  a  special scale-invariant form 
    and  will  vanish
    %v3 
     (modulo delta-function terms  with measure-zero support)
     due to the constraints of the massless scalar   kinematics. 
  
 As we shall  show   in section \ref{sec4} this  vanishing of the  four-scalar amplitude is, in fact, implied 
 by  the  global CHS symmetry   of the model. 
% and  demonstrate   that it implies   the  vanishing of the four-scalar amplitude.
 This  will 
 thus  justify our choice of the summation over spins prescription. 
 
  In section \ref{sec5} we will  consider  the one-loop amplitude given by  Fig.\ref{fig2}  and similar diagrams 
  limiting  the  computation  to  the local  UV divergent 
  $(\chh^* \chh)^2$  contribution  to  it. 
 Some concluding remarks will be made   in section \ref{Concl}.

  %old% 
% In  Appendix A we will discuss an alternative  regularisation using $\zeta$-function 
% leading to the same result as the prescriptions  described in section 3.2.
In  Appendix A we  will review the global CHS symmetry transformations. 
In Appendix B we will   present the explicit form of the cubic  and quartic vertices 
in the CHS   action  relevant for the computations in section 5. 
The transverse  traceless   gauge fixing and the  corresponding ghost action will 
be discussed in Appendix C.

%%%%%%%%%%%%%%%%%%%%%%%%%%%%%%%%%%%%%%%%%%%%%%%%%
\section{Scalar field  interacting with  conformal   higher  spin  fields}
\label{sec2}
%%%%%%%%%%%%%%%%%%%%%%%%%%%%%%%%%%%%%%%%%%%%%

Let us start   with a free   complex massless scalar
$\phi$  with the  flat  space   action 
\be 
	S_{\sst\rm free}[\phi]=\int d^{d}x\,\phi^{*}\,  
	\del^2 \,\phi\,.
	\label{2.1}
\ee
This  free theory   admits infinitely many  conserved (on-shell)  currents,
which are traceless due to  conformal invariance.
A generating function for  such traceless conserved currents may be  defined 
using an auxiliary  vector  $u_\m$ as  (see  \ci{Craigie:1983fb})
\be\lab{21}
	J(x,u)=\sum_{s=0}^{\infty}\frac1{s!}\,
	J^{\mu_{1}\cdots\mu_{s}} (x) \,u_{\mu_{1}}\cdots u_{\mu_{s}}  \ . 
\ee
Here 
\be\lab{31}
	J(x,u)=\Pi_{d}(u,\partial_x)\,\mathfrak{J}(x,u)\,, 
\ee
where $\mathfrak{J}(x,u)$ is the generating function of traceful currents
\be\lab{4}
	\qquad \mathfrak{J}(x,u)=\te \phi^{*}(x+{i\ov 2}\,u)\,\phi(x-{i\ov 2}\,u)\,,
\ee
and  $\Pi_{d}$ is an operator mapping the  traceful currents 
into traceless currents   \ci{Bekaert:2009ud,Bekaert:2010ky}\foot{Here  $(q)_n  = {\Gamma(q + n)\ov \Gamma(q)} $  is the Pochhammer symbol.}
\be
\label{25}
	\Pi_{d}(u,\partial_x)
	=\sum_{n=0}^{\infty}\frac1{n!\,(-u\cdot\partial_{u}-\frac{d-5}2)_{n}}
	\Big(\frac{u^{2}\,\partial_x^{2}-(u\cdot\partial_x)^{2}}{16}\Big)^{n}\,.
\ee
Let us consider an infinite set of couplings  of $\p$ to external  higher spin  fields  $h_s$ 
through these currents:
\be
	S_{\rm\sst int}[\phi,h]=\sum_{s=0}^{\infty} \frac1{s!}\,\int d^{d}x\, J^{\mu_{1}\cdots\mu_{s}}\,h_{\mu_{1}\cdots \mu_{s}}\,.
	\label{2.6}
\ee
Introducing 
 \be  \lab{299} 
 h(x,u) = \sum_{s=0}^{\infty}\frac1{s!}\,
	h_{\mu_{1}\cdots\mu_{s}} (x) \,u^{\mu_{1}}\cdots u^{\mu_{s}} \,, 
\ee
the coupling \rf{2.6}     may be written also as 
\be \lab{999}
	S_{\rm\sst int}[\phi,h] =\int d^{d}x\ {h}(x,\partial_{u})\,{J}(x,u)\,\big|_{u=0}\,. 
\ee
Due to the transversality and tracelessness of the currents  on the scalar  mass shell,   these couplings are invariant under 
\be
	\delta_{\sst\rm lin}\,h_{\mu_{1}\cdots\mu_{s}}
	=\partial_{(\mu_{1}}\varepsilon_{\mu_{2}\cdots\mu_{s})}
	+\eta_{(\mu_{1}\mu_{2}}\,\alpha_{\mu_{3}\cdots\mu_{s})}\ ,
	\label{2999}
\ee
provided   $\p$ is subject to its free equations   of motion. 
These are linearised conformal higher spin  (CHS) transformations \ci{Fradkin:1985am}. 
Off the scalar mass 
 shell, these symmetries are deformed to the  nonlinear  CHS ones   \ci{Segal:2002gd,Bekaert:2010ky}
generalising  the diffeomorphism and  Weyl transformations of the   Weyl   gravity
\be
	\delta_{\sst\rm CHS}\, h_{\mu_{1}\cdots\mu_{s}}
	=\delta_{\sst\rm lin}\, h_{\mu_{1}\cdots\mu_{s}}+
	\mathcal{O}(h)\,,
	\qquad\qquad 
	\delta_{\sst\rm CHS}\, \p=\mathcal{O}(\p)\,.
	\label{2.8}
\ee
For  \mt{s=0}   the field $h_0$ is a scalar coupled to $J_0 =   \phi^*   \phi$,  
for \mt{s=1}    we get a coupling of a vector $h_\m$  to $U(1)$ current, and for \mt{s=2} we get linearised metric $h_{\m\n}$ coupled to
energy-momentum  tensor.\foot{Other standard scalar coupling terms  such as 
$h_\m h^\m \phi^*\,\phi$ for electrodynamics
and ${(d-2)\ov 4 (d-1) }R\,\phi^*\,\phi$ for Weyl gravity 
can be absorbed  into a redefinition of $h_0$\,.}
The higher spin couplings are natural generalisations of these lower spin couplings.

%So far, higher spin fields $h_{s}$ were treated as background fields. 
%In general, one may add also their dynamics part 
Next, we may  supplement  \mt{S_{\rm\sst free}[\p]+S_{\rm\sst int}[\p,h]}
with the dynamical action for CHS fields $h_s$. 
 The functional of $h_{s}$  invariant under \rf{2.8}
 can be identified   with the local UV divergent part of the  induced action 
 found by integrating out some   number of additional 
 scalars  \ci{Segal:2002gd,Bekaert:2010ky}.  
 The  induced action  (discussed already  in the Introduction, see \rf{13},\rf{13a},\rf{14a})  
 may be written as \ci{Bekaert:2010ky}
\be
	%S_{\rm\sst ind}[h]=
	\Gamma[h]=\int d^{d}p\,  k(p) \, 
	\left(p^{2}\right)^{\frac{d-4}2}\,G(X,Y)\,\tilde h(p,u_{1})\,\tilde h(-p,u_{2})\,\Big|_{u_{i}=0}+
	\mathcal{O}(h^{3})\,,
	\label{300}
\ee
where $\tilde h(p,u)$  is the Fourier transform of $h(x,u)$ in \rf{299}  and  $k(p)$ is a spin-independent 
function 
\be   \lab{30}   k(p)  = c_1 \log { p^2\ov \Lambda^2}  + c_2  \  . \ee
 $\Lambda$ is a  UV cutoff (we omit power divergences)   and $c_1,c_2$  are simple numerical constants. 
The  operator  $G(X,Y)$   acting on $u_1,u_2$ is given by
\be\lab{26}
	G(X,Y) =
	\sum_{s=0}^{\infty}\frac{\Gamma(\frac{d-3}2)}{2^{4s}\,\Gamma(s+\frac{d-3}2)\,
	\Gamma(s+\frac{d-1}2)}\,C^{(\frac{d-3}2)}_{s} \Big(\frac{X}{\sqrt{Y}}\Big) \, Y^{\frac s2}\,,
\ee
where  $ C^{(\l)}_s (z)$ is the Gegenbauer polynomial   and  $X$ and $Y$ are  differential operators defined by 
\ba\lab{27}
	X\eq p^{2}\,\partial_{u_{1}}\!\cdot\partial_{u_{2}}
	-p\cdot\partial_{u_{1}}\,p\cdot\partial_{u_{2}}\,,\nn
	Y\eq \left[(p\cdot\partial_{u_{1}})^{2}-p^{2}\,\partial_{u_{1}}^{2}\right]
	\left[(p\cdot\partial_{u_{2}})^{2}-p^{2}\,\partial_{u_{2}}^{2}\right].
\ea
Keeping only the singular $\log \Lambda$ part  of  $k(p)$ or, equivalently, 
   replacing it  by a renormalized constant  $\ko= c_1 \log \mu^2$
(proportional to the number $N$  of  scalars  that were integrated out  and 
playing  the role of  the overall  inverse coupling constant)
we may define  the local  CHS action as 
%\item $h_s$ fields   are   described  just by the local  invariant part   of  the induced action \rf{300} 
 %(with $\ko$  
 \be
	S_{\rm\sst CHS}[h]= \ko  \int d^{d}p\,  \, 
	\left(p^{2}\right)^{\frac{d-4}2}\,G(X,Y)\,\tilde h(p,u_{1})\,\tilde h(-p,u_{2})\,\Big|_{u_{i}=0}+
	\mathcal{O}(h^{3})\,.
	\label{301}
\ee
The quadratic part of \rf{301}   represents a 
  collection of free conformal spin $s$ actions \ci{Fradkin:1985am}  
\be 
	\lab{1111}    
 	S_{\sst\rm CHS,2}[h_{s}] \sim \k  \int d^d x \ h_s \,\rP_s \,\del^{2s + d-4} \,h_s\,,  
\ee 
where as in \rf{14a}  the operator  $\rP_s$ is transverse traceless projector.
 $S_{\rm\sst  CHS,2}[h_{s}]$ is invariant under \rf{2999}
 and in $d=4$  may  be interpreted as the square of  the linearised    spin $s$   analog of 
Weyl tensor.
The important point here is that the relative  normalisation of 
 conformal spin $s$ fields  in the induced action are fixed %in $S_{\sst\rm ind}[h]$
 by  the coupling $S_{\rm\sst int}[\p,h]$ \eqref{2.6}
(other choices of normalisation 
 would  break the CHS symmetries \eqref{2.8}).\foot{One can also compute the $h^3$ term in the  local  CHS-invariant  $\log \Lambda$ part  of the induced action 
  \ci{Segal:2002gd,Bekaert:2010ky}. Extending  the construction of the non-linear   CHS action to  higher  orders in $h_s$   appears to be technically non-trivial 
  and may require  a new method  which is  non-perturbative in number of fields
  (see  in this connection  discussions of  the unfolding program for CHS fields \cite{Vasiliev:2001zy,Vasiliev:2007yc,Shaynkman:2014fqa}).} 
 
\iffa 
 In general, we   could  consider   both options:
 \begin{enumerate}
 \item $h_s$ fields   are   described  just by the local  invariant part   of  the induced action \rf{300} 
 (with $\ko$ playing  the role of  the overall  inverse coupling constant) 
 \be
	S_{\rm\sst CHS}[h]= \ko  \int d^{d}p\,  \, 
	\left(p^{2}\right)^{\frac{d-4}2}\,G(X,Y)\,\tilde h(p,u_{1})\,\tilde h(-p,u_{2})\,\Big|_{u_{i}=0}+
	\mathcal{O}(h^{3})\,.
	\label{301}
\ee
\item   $h_s$   dynamics is a non-local induced one  described by \rf{300},\rf{30}. 
\end{enumerate}
To  leading  quadratic order in $h_s$\,,  the difference   will only be  in a  universal modification of the kinetic term (and thus of the 
propagator)  $\ko \to k(p)$  under the integral (cf. \rf{30}). 

In what follows   we shall only  focus on the first case \rf{301}, i.e.  the local CHS dynamics.
 \fi

%%%%%%%%%%%%%%%%%%%%%%%%%%%%%%%%%%%%%%%%%%%%
\section{Four-scalar tree-level scattering  amplitude}
\label{sec3}
%%%%%%%%%%%%%%%%%%%%%%%%%%

Given the system of   CHS fields  coupled to a  free scalar   via \rf{2.6}, 
we can  study  the simplest four-scalar scattering process 
 with  the exchange of   all CHS fields (Fig.\ref{fig1}).
 This provides  an interesting example   when the issue of  definition of the sum over all spins   becomes important. 
In  \cite{Bekaert:2009ud}  a similar process  was analysed 
where the exchanged particles  were  the  standard  massless Fronsdal  higher spin  ones. 
There, the scattering amplitude was  obtained as a function of  infinitely many   undetermined 
 coupling constants between the massless  higher spin  fields and 
a scalar.
In the  present    case all the \mt{\phi\!-\!\phi\!-\!h_s} coupling constants
are fixed up to  an overall   factor   (the coupling constant  $\ko^{-1}$ of  the 
 CHS theory)  and   % that is the unique coupling constant of the theory.
as a result  the amplitude will be given by an 
explicit  expression  in terms of  a sum over  spins.

%%%%%%%%%%%%%%%%%%%%%%%%%%
\subsection{Conformal spin $s$ exchange} 
%%%%%%%%%%%%%%%%%%%%%%%%%%%%%%%%%%

To compute the relevant four-scalar amplitude we   start with the  vertex \rf{2.6} 
and consider integrating over $h_s$ (in  quadratic approximation  only)  while  keeping $\phi$  as external fields: 
\ba
	&&\left\la S_{\rm\sst int}[ \phi,h]\,S_{\rm\sst int}[ \phi,h]\,\right\ra_{\rm\sst 0} \nn
	&&=\,\sum_{s=0}^{\infty} \int \frac{d^{d}p}{(2\pi)^{d}}\,\frac 1{(s!)^{2}}\,
	\tilde J^{\mu_{1}\cdots\mu_{s}}(p)
	\left\la \tilde h_{\mu_{1}\cdots \mu_{s}}(p)\,\tilde h_{\nu_{1}\cdots\nu_{s}}(-p)\right\ra_{\rm\sst 0}
	\tilde J^{\nu_{1}\cdots\nu_{s}}(-p)\,. 
	\label{SS}
\ea
 Here  $\tilde J_s $  are the  Fourier transforms  of the bilinear conserved currents  in \rf{21}  and 
  the free   propagators   of the CHS fields   are   (in transverse traceless gauge) 
\be\lab{33}
	\left\la \tilde h_{\mu_{1}\cdots \mu_{s}}(p)\,\tilde h^{\nu_{1}\cdots\nu_{s}}(-p)\right\ra_{\rm\sst 0}
	= \frac{n_{s}}{2\k\ s!} 
	\frac{\rP^{\nu_{1}\cdots\nu_{s}}_{\mu_{1}\cdots \mu_{s}}(p)}{\left(p^{2}\right)^{s+\frac{d-4}2}}\, ,
\ee
where  \mt{\rP^{\nu_{1}\cdots\nu_{s}}_{\mu_{1}\cdots \mu_{s}}(p) = \delta^{\nu_{1}\cdots\nu_{s}}_{\mu_{1}\cdots \mu_{s}} + \ldots} 
is the   projector to transverse traceless totally symmetric tensors  and  $\ko$ is the overall coefficient  in \rf{301}.
Since the  propagators  are contracted with traceless and conserved
currents (the external scalar legs are assumed to be on-shell), 
all  other  terms denoted by dots in $\rP_s$     will drop out.  

The coefficients $n_{s}$ in \rf{33} are given by the normalisation of the quadratic part  in  \eqref{301}.
That they are completely fixed  is equivalent to  the fact 
 that  the \mt{\phi\!-\!\phi\!-\!h_s} coupling constants are all fixed.
%For the moment, let us keep $n_{s}$ unfixed and proceed the calculation.
Explicitly,  eq. \eqref{301} 
 contains different tensor structures represented
by different monomials in $X$ and $Y$. As we have remarked before, since the propagators
are contracted with traceless conserved currents, only traceless and transverse terms  are relevant. 
The $Y$ operator  contains  at least one trace or divergence, so
it is sufficient to  consider only  the $Y$-independent part of the  CHS action, i.e. to expand $G(X,Y)$  in \rf{26}  as 
\ba
\lab{42}
	G(X,Y)=\sum_{s=0}^{\infty}\frac{\Gamma(\frac{d-3}2)}{2^{3s}\,\Gamma(s+\frac{d-1}2)}\,\frac{X^{s}}{s!}
	+\mathcal O(Y)\,.
\ea
As a result, one   finds   
\be\lab{43}
	n_{s}=\frac{2^{3s}\,\Gamma\big(s+\frac{d-1}2\big)}{\Gamma(\frac{d-3}2)}\,.
\ee 	
Let us represent  \eqref{SS} as a sum over spins 
\be\lab{34}
	\left\la S_{\rm\sst int}[ \phi,h]\,S_{\rm\sst int}[ \phi,h]\,\right\ra_{\rm\sst 0}
	= \ko^{-1} \sum_{s=0}^{\infty} n_{s}\,V_{s}\,,
\ee
where the spin $s$ contribution  is   found to be 
\ba\lab{35}
	V_{s}\eq  \frac{1}{2\,s!} \int \frac{d^{d}p}{(2\pi)^{d}}\,\,\tilde J^{\mu_{1}\cdots\mu_{s}}(p)\,
	\frac{1}{(p^{2})^{s+\frac{d-4}2}}\,\tilde J_{\mu_{1}\cdots\mu_{s}}(-p)\nn
	\eq \frac{1}{2\,s!}
	\int \frac{d^{d}p}{(2\pi)^{d}}\,\frac{1}{(p^{2})^{s+\frac{d-4}2}}\,
	{(\partial_{u_{1}}\!\cdot\partial_{u_{2}})^{s}}\,
	 \Pi_{d}(u_1,i\,p)\,
	\tilde{\mathfrak{J}}(p,u_{1})\,\tilde{\mathfrak{J}}(-p,u_{2})\,\Big|_{u_{i}=0}\,,
\ea
where $\Pi_d$  was defined   in \rf{25}. 
The Fourier transform of the traceful-current generating function \rf{4}   is given by 
\ba\lab{36}
	\tilde{\mathfrak{J}}(p,u)\eq\int d^{d}x\, e^{-i\,x\cdot p}\,  
	{\te  \phi^{*}(x+{i\ov 2} \,u)\,\phi(x-{ i\ov 2} \,u)} \nn
	\eq\int \frac{d^{d}k\,d^{d}\ell}{(2\pi)^{2d}}\ 
	\tilde{\phi}^{*}(k)\,\tilde{\phi}(\ell)\,e^{u\cdot \frac{k+\ell}2}
	(2\pi)^{d}\,\delta^{(d)}(p+k-\ell)\,.
\ea
Using this expression  %and factoring all $s$-independent part,
we  can represent  $V_{s}$ in \rf{34} as 
\ba\lab{37}
	V_{s}\eq \frac{1}{2}\int \frac{d^{d}k_{1}\,d^{d}\ell_{1}\,d^{d}k_{2}\,d^{d}\ell_{2}}{(2\pi)^{4d}}\,
	(2\pi)^{d}\,\delta^{(d)}(k_{1}+k_{2}-\ell_{1}-\ell_{2}) \nn
	&& \qquad
	\times\,\tilde{\phi}^{*}(k_{1})\,\tilde{\phi}(\ell_{1})\,
	\tilde{\phi}^{*}(k_{2})\,\tilde{\phi}(\ell_{2})\,
	A_{s}(k_{1},k_{2},\ell_{1},\ell_{2})\,,
\ea
where $A_{s}$  is    the spin-$s$ exchange amplitude  $({p=k_{1}-\ell_{1}=\ell_{2}-k_{2}})$
\be\lab{38}
	A_{s}(k_{1},k_{2},\ell_{1},\ell_{2})=\frac{1}{2\,(p^{2})^{s+\frac{d-4}2}}\,
	\frac{(\partial_{u_{1}}\!\cdot\partial_{u_{2}})^{s}}{s!}\,
	\Pi_{d}(u_1,i\,p)\,e^{\ha\left[u_{1}\cdot (k_{1}+\ell_{1})+
	u_{2}\cdot (k_{2}+\ell_{2})\right]}\,\Big|_{u_{i}=0}\,.
\ee
Using the explicit expression for  $\Pi_{d}$ in \rf{25} 
the resulting  $\mathsf t$-channel  amplitude  due to spin $s$ exchange 
is  found to be 
\ba\lab{39}
	A^{(\mathsf t)}_{s}(\mathsf s,\mathsf t, \mathsf u)\eq\frac{1}{2(-4)^{s}\,(-\mathsf t)^{\frac{d-4}2}}
	\sum_{n=0}^{[s/2]}\frac1{2^{2n}\,n!\,(s-2n)!\,(-s-\frac{d-5}2)_{n}}
	\left(\frac{\mathsf s-\mathsf u}
	{\mathsf s+\mathsf u}\right)^{s-2n}\nn
	\eq
	\frac{1}{2(-8)^{s}\,(\frac{d-3}2)_{s}}\,\frac{1}{(-\mathsf t)^{\frac{d-4}2}}\,
	C^{(\frac{d-3}2)}_{s}\left(\frac{\mathsf s-\mathsf u}
	{\mathsf s+\mathsf u}\right). 
\ea
Here  $\mathsf s,\mathsf t, \mathsf u$ are the Mandelstam variables
 (with $ \mathsf s+ \mathsf t +  \mathsf u=0$ in the present massless  scalar case)
 and $C^{(\lambda)}_{n}(z)$ is  the  Gegenbauer polynomial.

 Since the  theory under consideration is  conformal,  the amplitude has a manifestly  scale-covariant form. 
 In particular, in  \mt{d=4}   it depends only on ratio of the Mandelstam variables  
 (also, in \mt{d=4} the Gegenbauer polynomial reduces
  to the Legendre one).
  \iffa
  \foot{Let us note that in the case when the exchanged   higher spin fields are described 
 by the non-local induced action \rf{300}, \rf{30}  the amplitude in \rf{38}, \rf{39} 
 gets  just an  extra overall  momentum dependent factor  $1/k(p)$    in the  denominator, i.e. 
 \rf{39}   gets  a factor $[\log  ( a\,\mathsf t) ]^{-1}$    where $a$ is  a constant  depending on renormalisation scale. 
 This  would   not  affect the  issue of summation over spins  discussed  below.}
 \fi
 
The total summed over spins 
$\mathsf t$-channel amplitude is thus  given by (cf. \rf{34}, \rf{37})
\be\lab{40}
	A^{(\mathsf t)}
	(\mathsf s,\mathsf t, \mathsf u)  =\ko^{-1} \sum_{s=0}^{\infty} n_{s}\,   A^{(\mathsf t)}_{s}(\mathsf s,\mathsf t, \mathsf u)
	= \ko^{-1} \,\frac{1}{2\,(-\mathsf t)^{\frac{d-4}2}}\,F_{d}\Big(-\frac{\mathsf s-\mathsf u}
	{\mathsf s+\mathsf u}\Big)\  ,
\ee
where the function $F_{d}(z)$ is given by 
\be\lab{41}
	F_{d}(z)=\sum_{s=0}^{\infty}\frac{n_{s}}{2^{3s}\,(\frac{d-3}2)_{s}}\,
	C^{(\frac{d-3}2)}_{s}(z)\,.
\ee
Using        the expression for $n_s$ in \rf{43}\,,    $F_{d}(z)$    simplifies to 
\be
	F_{d}(z)=\sum_{s=0}^{\infty}\left(s+\alpha_{d}\right) C^{(\alpha_{d})}_{s}(z)\,,
	\qquad\qquad 
	\te \alpha_{d}\equiv \frac{d-3}2\,.
	\label{313}
\ee
For generic values of $z$\,, the  sum over spins diverges  and thus needs to be   defined 
 with a  certain regularisation prescription.

%%%%%%%%%%%%%%%%%%%%%%%%%%
\subsection{Summing  over  spins} 
%%%%%%%%%%%%%%%%%%%%%%%%%%%%%%%%%%

In general,  a particular   definition of the sum over spins  and thus the resulting expressions 
for the   scattering amplitudes   should be  consistent with  the underlying symmetries of the theory.\foot{One may  draw an 
analogy with the Veneziano amplitude   in string theory where one  also sums over an infinite number of  different 
(massive) field    contributions. When computing it in string field theory context, one would also need to choose 
 a particular   summation over modes   prescription. This prescription  is  selected  automatically  in  the first-quantised world sheet approach 
 in which the 2d conformal invariance and the associated space-time symmetries  are  built in.}
We shall   return to this point below but let us   first proceed   formally, choosing a natural cutoff prescription  to define the sum over $s$. 
Let us introduce a  parameter $w= e^{-\varepsilon}  < 1 $ (with $ \varepsilon \to 0$),  compute the sum  
 and  then  define   \eqref{313}  as a  limit $w\to 1$ 
\be
F_{d}(z)=\lim_{w\to1}\, F_{d}(z,w)\ , \ \ \ \ \ \ \ \ \ \ \ \ \ 
	F_{d}(z,w)=\sum_{s=0}^{\infty}\left(s+\alpha_{d}\right)\,w^{s}\,C^{(\alpha_{d})}_{s}(z)\,.
	\label{314}
\ee
We may   write $F_{d}(z,w)$   as 
\be\lab{45}
	F_{d}(z,w)=w^{1-\alpha_{d}} \frac{d}{dw}\Big(
	w^{\alpha_{d}} \sum_{s=0}^{\infty}w^{s}\,C^{(\alpha_{d})}_{s}(z)\Big) \ ,
\ee
and use  the expression  for  the generating function $\sum_{s=0}^{\infty}w^{s}\,C^{(\alpha_d)}_{s}(z)
= (1- 2 z w + w^2)^{-\alpha_d}$
  for the Gegenbauer polynomials to 
define the regularized   expression for $F_{d}(z,w)$   by an analytic continuation:\foot{The radius of convergence of the series
in $w$  is not greater than 1
(it  is 1 when $|z|< 1$  and  $e^{-x}$ when $|z|=\cosh x \ge 1$)  
so  the direct evaluation of $F_{d}(z,1)$ gives a divergent expression.}
\be
\lab{316} 
	F^{\sst\rm reg}_{d}(z,w)=\alpha_{d}\,\frac{1-w^{2}}{(1-2z\,w+w^{2})^{\alpha_{d}+1}}\,.
\ee
 Notice that $F^{\sst\rm reg}_{d}(z,1)$ happens to vanish  for $z\neq1$,
 while for $z=1$\,, we get
\be\lab{46}
	F^{\sst\rm reg}_{d}(1,w)=\alpha_{d}\,\frac{1+w}{(1-w)^{d-2}}\,,
\ee
which diverges as $w\to 1$\,. 
Thus $F^{\sst\rm reg}_{d}(z)$ is a  particular distribution with  support 
localised at $z=1$. In fact, it  is just proportional to  the $(d-4)$-th derivative
of the delta-function, i.e.\foot{Starting with \rf{316}
and 
 changing the  variables \mt{z=x+w}, \  \mt{\e^{2}=1-w^{2}} we  get
\mt{F_{d}^{\rm reg}(x,\e)=\a_{d}\,\frac{\e^{2}}{(x^{2}+\e^{2})^{\a_{d}+1}}}. 
As a  result, 
\mt{F_{d}^{\rm reg}(z)= \lim_{\e \to 0} F_{d}^{\rm reg}(x,\e)=\frac{(-1)^{d-4}}{(d-4)!}\,\delta^{[d-4]}(x)\,.}
}
\be\lab{47}
	F^{\sst\rm reg}_{d}(z)=\frac{(-1)^{d-4}}{(d-4)!}\,\delta^{[d-4]}(z-1)\,, \qquad  \ \ \ {\rm i.e.} \qquad  \ \ \ \ \ \ 
	F^{\sst\rm reg}_{4}(z)=\delta(z-1) \ .
\ee
%In $d=4$   this becomes simply $\delta(z)$. 
The   above  regularisation of the sum over spins   is  essentially the same as  the one  used in 
\ci{Giombi:2014iua,Beccaria:2014jxa,Beccaria:2015vaa}  in the context   of  higher spin partition functions.
In the case of  CHS   theory in $d$ dimensions (or $d$-dimensional  boundary theory) 
the sum $\sum_{s=0}^\infty   f_d (s)$ was  first replaced     by  the convergent sum 
$\sum_{s=0}^\infty   e^{-  \varepsilon ( s + \alpha_d) }  f_d(s)$
where $\alpha_d = { d-3 \ov 2}$  and then
after  taking the limit $\varepsilon \to 0$ all $ 1\ov \varepsilon^n$ poles  were dropped.

The same result  \rf{47}  
is found also using another natural regularisation  prescription utilizing  integral representation for the Gegenbauer 
%AT
polynomials.
% (yet another regularisation using $\zeta$-function is discussed in Appendix A). 
For simplicity, let us focus on the $d=4$ case where \rf{313}  reduces to
\be\lab{48}
	F_{4}(z)=\sum_{s=0}^{\infty}\big(s+\te \frac12\big)\ P_{s}(z)\,.  
\ee
Here   $P_s=  C^{(1/2)}_s$   is the  Legendre polynomial.
 The  idea  
is to use    the integral   representation
\be\lab{49}
	P_{s}(z)
	=\frac1{\pi}\int_{0}^{\pi}dx\left(z+\sqrt{z^{2}-1}\,\cos x\right)^{s}\,, 
\ee
and interchange the  summation over $s$  with  the  integration.   
 Performing  first  the sum  we  find  the following integrand 
\be\lab{50}
	\sum_{s=0}^{\infty}\big(s+\frac12\big)
	\left(z+\sqrt{z^{2}-1}\,\cos x\right)^{s}
	=\frac{z+1+\sqrt{z^{2}-1}\,\cos x}{2\,(z-1+\sqrt{z^{2}-1}\,\cos x)^{2}}\,.
\ee   
Here we have also used an analytic continuation 
since for any $x\in[0,\pi]$\,, there exists  such $z$  that the series is divergent.
Performing  the $x$-integral  we  get
\be\lab{51}
	F^{\rm reg}_{4}(z)=\frac1{\pi}\int_{0}^{\pi}dx\,
	\frac{z+1+\sqrt{z^{2}-1}\,\cos x}{2\,(z-1+\sqrt{z^{2}-1}\,\cos x)^{2}}=\delta(z-1)\,,
\ee
i.e. the same result as in  \rf{47}.

%%%%%%%%%%%%%%%%%%%%%%%%%%%%%%
\def \phis {\phi^*}

\subsection{Total amplitude in $d=4$}
%%%%%%%%%%%%%%%%%%%%%%

In  the case of a   complex scalar scattering  
\mt{\phi\,\phi\,\to\, \phis\, \phis}  in $d=4$ 
one  finds the total amplitude by  adding   the $\mathsf t$-channel and the $\mathsf u$-channel 
contributions following from \rf{40} and \rf{47}, \rf{51} 
\be\lab{60}
	A_{\phi\phi\to\phis\phis}=\frac{\ko^{-1} }4
	\left[\delta\left(\frac{\mathsf s}{\mathsf t}\right)+\delta\left(\frac{\mathsf s}{\mathsf u}\right)
	\right] \ . 
\ee
This   unfamiliarly  looking amplitude 
  actually vanishes for  physical momenta   due to massless   kinematics. 
Indeed,  choosing  the c.o.m. frame (\mt{\vec p_1  + \vec p_2 =0= \vec p_3 + \vec p_4  })
and   introducing the scattering angle $\theta$  for which 
\mt{\cos \theta = {\vec p_1 \cdot  {\vec p_3} \ov |\vec p_1 | | \vec p_3 | }}
one can show   (using $E_i =  |\vec p_i |$) that
${\mathsf s\ov \mathsf t }= -{1\ov  \sin^{2}{\tfrac{\theta}2} } , \ \   
{\mathsf s \ov \mathsf u} = - {1\ov \cos^{2}{\tfrac{\theta}2} }.$\foot{In general, there  may  be  a possible subtlety in the  collinear limit when $p^\mu_1 =  r p^\m_2$ 
and one cannot go to the  c.o.m. frame   but this  limit  requires complex   momenta  and its  significance in the present context 
 is unclear.}
Thus   the arguments  of the delta-functions   never  vanish for real $\theta$, i.e. we get 
%v3
\be\lab{61}
	A_{\phi\phi\to\phis\phis}=0\,.
\ee
For the $\phi\, \phi^* \,\to\, \phi\, \phi^*$ scattering, we  find
\be\lab{62}
	A_{\phi\phi^{*}\to\phi\phi^{*}}=\frac{\ko^{-1}}4
	\left[\delta\left(\frac{\mathsf u}{\mathsf t}\right)
	+\delta\left(\frac{\mathsf u}{\mathsf s}\right)\right]=
	\frac{\ko^{-1}}4\Big[\delta\left(\cot^{2}\tfrac{\theta}2\right) + \delta\left(\cos^{2}\tfrac{\theta}2\right)\Big]\,,
\ee
where  the two delta-functions  correspond to the $\mathsf t$-channel and the $\mathsf s$-channel 
contributions, respectively. 
%v3
%These  two contributions cancel each other,  so that   again  the total amplitude vanishes 
Here   the arguments of the  delta-functions   do not vanish unless $\theta= \pi$  so that excluding this special point  we get 
\be\lab{63}
	A_{\phi\phi^*\to\phi\phi^*}=0\,.
\ee
One may also consider the real scalar case  when only the even spin currents in \rf{21}   are non-vanishing 
and thus only the even spin  CHS exchanges are   contributing. Then only 
 the even $z$ part of the  function in   \rf{41}, \rf{51}    is relevant  and we get 
 for the total  four-scalar   scattering amplitude
 %v3
\ba\lab{64}
	&&A_{\phi\phi\to\phi\phi}^{\sst(\mathbb R)}=\frac{\ko^{-1}}8
	\left[
	\delta\left(\frac{\mathsf u}{\mathsf t}\right) + \delta\left(\frac{\mathsf s}{\mathsf t}\right)
	+\delta\left(\frac{\mathsf u}{\mathsf s}\right)
	+\delta\left(\frac{\mathsf t}{\mathsf s}\right)
	+\delta\left(\frac{\mathsf t}{\mathsf u}\right)
	+\delta\left(\frac{\mathsf s}{\mathsf u}\right)
	\right]\\
	&&=\,
	\frac{\ko^{-1}}8
	\Big[
	\delta\left(\cot^{2}\tfrac{\theta}2\right) 
	+\delta\left(\csc^{2}\tfrac{\theta}2\right)
	+\delta\left(\cos^{2}\tfrac{\theta}2\right)
	+\delta\left(\sin^{2}\tfrac{\theta}2\right)
	+\delta\left(\tan^{2}\tfrac{\theta}2\right)
	+\delta\left(\sec^{2}\tfrac{\theta}2\right)
	\Big]\,.\nonumber
\ea
Here the first two delta-functions  come from the $\mathsf t$-channel,
the middle two   from the $\mathsf s$-channel
and the last two from the $\mathsf u$-channel  exchange. 
%v3
%The contributions from the three channels cancel against each other and we get again 
Here the  arguments   of the delta-functions do not vanish unless $\theta=0, \pi$  and thus excluding these points we get 
\be\lab{65}
	A_{\phi\phi\to\phi\phi}^{\sst(\mathbb R)}=0\,.
\ee
To conclude, while the individual spin $s$ exchange   contributions are nontrivial, the total 
amplitude   vanishes if computed   with a particular prescription for  summation over spins. 
As we shall argue   below, the vanishing of the  four-scalar scattering amplitude 
 is actually  implied   by    the global  CHS symmetry of the  theory.

%%%%%%%%%%%%%%%%%%%%%%%%%%%%%%%%%%%%%%%%%%%

\section{Constraints 
 of  conformal higher spin symmetry  on   scalar  amplitudes}
\label{sec4}
%%%%%%%%%%%%%%%%%%%%%%%%%

We have seen that the tree-level scattering amplitude 
 vanishes when a  particular regularisation  is used 
 to define  the  summation over all  exchanged  spins. 
The   principle   that should   be selecting one regularization over  the other   should be 
the preservation of underlying symmetries  of the 
theory.\foot{One  possible analogy  is  with summation over the  Kaluza-Klein modes in  a 5d  theory compactified on a circle. 
Viewed as a 4d theory it involves sum over an  infinite number of  KK   mode  contributions 
with manifest symmetry being only 4d Lorentz symmetry, but the requirement  of preservation of the original 5d Lorentz symmetry 
should  impose constraints on how one should perform the 
sum  to recover the result found directly in 5d.}

The system of  CHS   fields  coupled to  massless  scalar  has 
the {\it global} CHS symmetry which plays an analogous role  to  Lorentz or conformal 
symmetry in   standard    field theory.
One    may thus  require  the
consistency of  a prescription  of summation over spins with this symmetry.  
For example,  the  introduction of the regularization factor $w^s $  in \rf{314} may be  implemented by  adding it to the CHS propagator   in \rf{33}.
This  translates into  the following   modification   of  the  quadratic part of the CHS action \rf{301} (see \rf{26},\rf{27}) 
\be\lab{411}
	S^{\rm reg}_{\rm\sst CHS, 2}[h;\o]=\int d^{d}p\, 
	\left(p^{2}\right)^{\frac{d-4}2}\,G(\o^{-1}\,X,\o^{-2}\,Y)\,\tilde h(p,u_{1})\,\tilde h(-p,u_{2})\,\Big|_{u_{i}=0}\, .
\ee
One   may then ask if   this regularized action still preserves the  global CHS symmetry  which is reviewed in Appendix \ref{appsym}.

Below  we   will  demonstrate that the vanishing of the tree amplitude  found in the previous section is  actually 
implied   by the invariance  under  a  particular   subset of  global CHS symmetry transformations. 
 This provides an  evidence of  a consistency of the regularization of the sum over spins 
used in Section \ref{sec3}. 

%%%%%%%%%%%%%%%%%%%%%%%%%%%%%%%%%%%%%%%%%%%%%%%%%%%%%

%%%%%%%%%%%%%%%%%%%%%%%%%%%%%%%%%%%%%%%%%%%%
%\subsection*{Constraints of CHS symmetry  on   scalar  amplitudes}
%%%%%%%%%%%%%%%%%%%%%%%%%%%%

Assuming   that CHS symmetry is free from anomalies,\foot{Possible 
anomalies  from loop graphs   may cancel if one sums  over all   CHS fields. 
 Indeed,  it was demonstrated  in \ci{Giombi:2013yva,Tseytlin:2013jya} 
that $a$-coefficient of Weyl anomaly of the \mt{d=4} CHS   theory  vanishes assuming
a particular prescription of summation over spins. 
The same  may apply  also  to the $c$-coefficient of 4d Weyl anomaly 
\ci{Tseytlin:2013jya,Giombi:2014iua,Beccaria:2014xda,Beccaria:2015vaa}.
As the Weyl   symmetry is one of the   CHS  gauge 
 symmetries, this is an indication that the same  may apply to all  algebraic  CHS symmetries. 
} 
we   would   like to  analyze how  the  global CHS symmetry  of the scalar  action coupled to the  CHS fields 
constrains the  correlators (and thus the scattering amplitudes)   of massless  scalar fields.
%no  tree level  as  we  induce CHS action 
The  global     CHS symmetry should constrain possible  interaction terms in the effective action
for the scalars  (with   CHS  fields  integrated out, i.e. appearing only on internal lines).  
In fact,  it   may  prohibit any non-trivial interaction terms, 
i.e.   may  imply the vanishing of the  corresponding S-matrix.

%Since the CHS symmetry contains the conformal algebra, the four-point functions of $N$ conformal scalars
%are subject to take the following form:
%\ba
%	&&\la \phi^{a_1}(x_{1})\,\phi^{a_2}(x_{2})\,\phi^{a_3}(x_{3})\,\phi^{a_4}(x_{4})\ra
%	=\frac{\eta^{a_{1}a_{3}}}{|x_{13}|^{d-2}}\,\frac{\eta^{a_{2}a_{4}}}{|x_{24}|^{d-2}}\,f(u,v)\nn
%	&& \quad +\,\frac{\eta^{a_{2}a_{3}}}{|x_{23}|^{d-2}}\,\frac{\eta^{a_{1}a_{4}}}{|x_{14}|^{d-2}}\, f(\frac uv,\frac 1v)
%	+\frac{\eta^{a_{1}a_{2}}}{|x_{12}|^{d-2}}\,\frac{\eta^{a_{3}a_{4}}}{|x_{34}|^{d-2}}\, f(\frac 1u,\frac vu)\Big]
%	+\mathcal O(\hbar)\,,
%\ea
%where $u$ and $v$ are the usual cross ratios:
%\be
%	u=\frac{|x_{12}|\,|x_{34}|}{|x_{13}|\,|x_{24}|}\,,
%	\qquad v=\frac{|x_{23}|\,|x_{14}|}{|x_{13}|\,|x_{24}|}\,.
%	\label{cross ratio}
%\ee
%$f$ is a symmetric function, $f(u,v)=f(v,u)$, which cannot determined using only conformal symmetry.
%Now, we use the CHS symmetry to constrain $f$\,. 

Among the  infinitely many  global CHS transformations \rf{488}, 
 let us  consider  
the ``hyper-translations"  (cf.  \rf{4133}):\footnote{Here we shall  ignore the  trace parts: 
 the trace parts of \eqref{4244} correspond to the trivial symmetries (vanishing on equations of motion) 
that  will not give any useful  conditions for  the correlators. 
There is no problem in including such symmetries  back   if needed. %  for technical simplifications.
}
\be
	\delta \phi(x)=\varepsilon^{\mu_1  ....\mu_r} \,\partial_{\mu_{1}}\cdots\,\partial_{\mu_{r}}\,\phi(x)\,.
	\label{4244}
\ee
Here $\varepsilon^{\mu_1  ....\mu_r} $ is a constant parameter. 
For simplicity, let us    restrict the discussion  to 
the case of real scalars, so that $r$ will  take  only odd values. 
Choosing     $\varepsilon^{\mu_1  ....\mu_r} $  proportional to a product 
$ 
 y^{\mu_1} ... y^{\mu_r}$    where   $y^{\mu}$  is an arbitrary vector  
we  conclude that  \rf{4244} implies  also the invariance under 
\be\lab{a2}
	\delta \phi(x)=(e^{y\cdot\partial_{x}}-e^{-y\cdot\partial_{x}})\phi(x)
	=\phi(x+y)-\phi(x-y)\,.
\ee
The invariance of the scalar four-point correlation 
 function under   such   symmetry implies
\ba
	&&\la \phi(x_{1}+y)\,\phi(x_{2})\,\phi(x_{3})\,\phi(x_{4})\ra
	+\la \phi(x_{1})\,\phi(x_{2}+y)\,\phi(x_{3})\,\phi(x_{4})\ra\nn
	&&+\,\la \phi(x_{1})\,\phi(x_{2})\,\phi(x_{3}+y)\,\phi(x_{4})\ra
	+\la \phi(x_{1})\,\phi(x_{2})\,\phi(x_{3})\,\phi(x_{4}+y)\ra\nn
	&&-\,(y\leftrightarrow -y)=0\,. \lab{a22}
\ea
Translated to the   momentum space this   constraint becomes 
\be\lab{a3}
	\sin(p_{12}\cdot y)\,\sin(p_{13}\cdot y)\,\sin(p_{14}\cdot y)\,
	\la \tilde\phi(p_{1})\,\tilde\phi(p_{2})\,
	\tilde\phi(p_{3})\,\tilde\phi(p_{4})\ra=0\,,
\ee
where $p_{ij}=\ha (p_{i}+p_{j})$ and we have used trigonometric identities and momentum conservation, $p_{1}+p_{2}+p_{3}+p_{4}=0$\,.
Making special choice of  the vector $y^\m$  as 
\be\lab{a5}
	y^\m= a\,p^\m_{12}+ b \,p^\m_{13}+ c \,p^\m_{14}\ , 
\ee
where  $a,b,c$ are some arbitrary parameters, and applying the  condition \rf{a3}
  to the case of the   on-shell  scattering amplitude of four real scalars  (cf. \rf{64}) 
we get    (using that  $p_i^2=0$)
\be\lab{a4}
	\sin(\fo a\,\mathsf s)\,\sin(\fo b\,\mathsf t)\,\sin(\fo c\,\mathsf u)\,
	A^{\sst(\mathbb R)}_{\phi\phi\to\phi\phi}(\mathsf s,\mathsf t,\mathsf u)=0\,.
\ee
Since $a,b,c$ are arbitrary, eq.\eqref{a4}  is equivalent to \ 
$\mathsf s\,\mathsf t\,\mathsf u\,A^{\sst(\mathbb R)}_{\phi\phi\to\phi\phi}=0$\,,
and its solution is given by the distribution,
\be\lab{a6}
	A^{\sst(\mathbb R)}_{\phi\phi\to\phi\phi}(\mathsf s,\mathsf t,\mathsf u)
	=k_1(\mathsf t,\mathsf u)\,\delta(\mathsf s)
	+k_2(\mathsf u,\mathsf s)\,\delta(\mathsf t)
	+k_3(\mathsf s,\mathsf t)\,\delta(\mathsf u)\,,
\ee
with a priori arbitrary functions $k_i$.
In addition, we may  use   also  the conformal symmetry  which   is   a   sub-algebra of the CHS symmetry. 
In particular, in $d=4$  the amplitude should   be invariant   under the dilatation symmetry (cf. \rf{64}), i.e. under  the 
rescaling  of momenta  by a  real constant $\l$  
\be\lab{a7}
	A^{\sst(\mathbb R)}_{\phi\phi\to\phi\phi}(\l^2\,\mathsf s,\l^2\,\mathsf t,\l^2\,\mathsf u)=
	A^{\sst(\mathbb R)}_{\phi\phi\to\phi\phi}(\mathsf s,\mathsf t,\mathsf u) \,. 
\ee
This condition restricts  $k_i$ to be  homogeneous functions of degree one.
%v33
Finally, we should impose the crossing symmetry condition, i.e. $k_i(x,y) = k(x,y)$. 
Using $\mathsf s+\mathsf t+\mathsf u=0$
(which is implied already by the on-shell conditions used to arrive at \rf{a4})
we have   essentially  unique  choice  for the  homogeneity-one function $k$.       Given that  under the delta-function 
 $\mathsf s$  in $k(\mathsf t,\mathsf  u) =k(\mathsf  t, -\mathsf  s -\mathsf t)$ can be set  to zero    
 %   = c_1  t  + c_2  u
 for  linear  function  we have 
$k(\mathsf t,\mathsf u)\delta(\mathsf  s)  \sim   \mathsf  t\,  \delta(\mathsf  s) $ 
 but this  is trivial upon symmetrization required by crossing symmetry. So   we are left only  with the modulus choice: 
    $k(\mathsf t,\mathsf u)  \delta(\mathsf s)  \sim   |\mathsf t | \delta(\mathsf  s)$. %  = \delta({s\ov t})$. 
Written in  symmetric  form  we   thus find the following  non-trivial solution
%Moreover, the crossing symmetry of the amplitude requires $k_i(x,y)=k(x+y)$\,.
%Combining all these  results, we obtain
\be \lab{a66}
	A^{\sst(\mathbb R)}_{\phi\phi\to\phi\phi}(\mathsf s,\mathsf t,\mathsf u)
	=c\Big[ (|\mathsf t|+|\mathsf u|)\,\delta(\mathsf s)
	+(|\mathsf u|+|\mathsf s|)\,\,\delta(\mathsf t)
	+(|\mathsf s|+|\mathsf t|)\,\delta(\mathsf u)\Big],
\ee
where $c$ is an arbitrary  overall constant. 
%v3
This is equivalent to  the  expression in \rf{64}  obtained  above  by the direct   computation of the scattering amplitude 
(note that $|\mathsf  t| \delta (\mathsf  s) = \delta({\mathsf  s\ov \mathsf t})$, etc.)\foot{We   thank M. Taronna  for  pointing out a mistake 
in the above  argument  in  an  earlier version of this paper.}
%  and observing that  another equivalent form  of \rf{a66} 
% is  $A^{\sst(\mathbb R)}_{\phi\phi\to\phi\phi}(\mathsf s,\mathsf t,\mathsf u)
%	= c\big[ |\mathsf t - \mathsf u|\,\delta(\mathsf s)
%	+|\mathsf u - \mathsf s|\,\,\delta(\mathsf t)
%	+|\mathsf s - \mathsf t|\,\delta(\mathsf u)\big] =2 k\big[ |\mathsf t |\,\delta(\mathsf s)
%	+|\mathsf u |\,\,\delta(\mathsf t)
%	+|\mathsf s |\,\delta(\mathsf u)\big] $
%since $|\mathsf t - \mathsf u|\,\delta(\mathsf s)= |2\mathsf t + \mathsf s|\,\delta(\mathsf s)
%=2 |\mathsf t |\,\delta(\mathsf s)$, etc.
and  thus  vanishes  for physical momenta apart from measure zero domain in phase 
%v33
 space.\foot{Because of $\mathsf s+\mathsf t+\mathsf u=0$   there are several 
 possible expressions that  reduce to \rf{a66}. 
  We can consider three different cases: 
  (i) none of $\mathsf s,\mathsf t, \mathsf u$  vanishes; (ii) only one of them  vanishes; (iii)   two of them (and thus also the third) 
  vanish. In the  first case the amplitude   is zero because of the delta-functions and in the   third case -- because of the prefactors.
   In the only non-trivial  second case 
  we get the expression equivalent to  \rf{a66} or \rf{64}.}

%Finally, imposing the condition $\mathsf s+\mathsf t+\mathsf u=0$\,,
%we conclude that the amplitude completely vanishes.
 
This  formal argument appears to apply  not only at the tree but also at the loop level 
if the global CHS symmetry is not anomalous. 
It should also apply to the complex scalar scattering case. 
%v3
As we have  already seen  in Section \ref{sec3},    the tree-level  scalar amplitude  
indeed vanishes (modulo delta-function terms)  in a particular regularization of the sum under spins 
 which should thus   be consistent with the CHS symmetry. 
 
It would    be interesting to  directly verify  this vanishing   also  for the full one-loop  
 on-shell scalar amplitude.  We shall  address the computation of the loop amplitude   in the  next section.

%%%%%%%%%%%%%%%%%%%%%%%%%%%%%%%%%%%%
\section{One-loop   corrections}
\label{sec5}
%%%%%%%%%%%%%%%%%%%%%%%%%%%%%

Let us  now turn attention to the  quantum corrections.  
Here we  will not    compute the full one-loop  correction to four-scalar 
 amplitude  (which is expected to vanish    in view of the symmetry  argument 
 in the previous section) 
  but  address only  the  question about UV singular part of the amplitude.
  We shall  consider the case of  dimension $d=4$. 
 
In 4d scalar QED,   the  four-scalar one-loop  amplitude  contains  logarithmic UV divergence 
coming from  loop diagrams  with  spin-one  propagators, and
 similar  divergences are expected  in  each  conformal higher spin  loop.
 One may ask   if these divergences may   go away   after one sums  over all  spins, i.e. if 
   four-scalar one-loop S-matrix is UV finite in the model  of massless 
 scalar  coupled to CHS theory.  Below  we shall address this question
 by explicitly calculating such UV divergence.

Since the only coupling  constant $\k$  in this theory \rf{2.1},\rf{2.6},\rf{301}   is dimensionless
on dimensional grounds the  only possible logarithmic  UV   divergence in the on-shell 
 effective action 
is  proportional to the  local  term $\int d^4 x\,(\phi^* \phi)^2$.
%local zero-momentum part of the scalar amplitude. 
In order to compute the coefficient of  this  term in the one-loop effective action  
it is sufficient  to consider  the  background  field $\phi$ to be constant,  
i.e. to assume that the  external  legs  in four-scalar one-loop  amplitude are taken at zero momentum (which is a particular on-shell 
point in a massless  scalar theory, so the  result should be gauge-independent). 
Henceforth we shall focus only  on the
amplitudes with vanishing external momenta.

% The full UV divergence proportional to  $\int  d^4 x \,  (\phi^* \phi)^2$
%receive contributions from
%different one-loop diagrams with different spin exchanges. 
% Let  us comment on the possible diagrams 
% whether their contributions are zero or non-zero in  zero-momentum limit.
\subsection{Diagrams  contributing to four-scalar scattering  amplitude}

\subsubsection*{Box diagram}

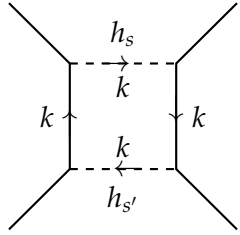
\begin{figure}[h]
\centering
\begin{tikzpicture}
\draw [thick] (-1.5,1.5) -- (-0.7,0.7) -- (-0.7,-0.7) -- (-1.5,-1.5);
\draw [thick] (1.5,1.5) -- (0.7,0.7) -- (0.7,-0.7) -- (1.5,-1.5);
\draw [thick,dashed] (-0.7,0.7) -- (0.7,0.7);
\draw [thick,dashed] (-0.7,-0.7) -- (0.7,-0.7);
\node at (-1,0) {$k$};
\node at (1,0) {$k$};
\node at (0,0.4) {$k$};
\node at (0,-0.4) {$k$};
\node at (0,1.1) {$h_{s}$};
\node at (0,-1.1) {$h_{s'}$};
\node at (0.7,0.1) {$ \downarrow$};
\node at (-0.7,-0.1) {$ \uparrow$};
\node at (-0.1,0.675) {$ \rightarrow$};
\node at (0.07,-0.725) {$ \leftarrow$};
\end{tikzpicture}
\caption{Box diagram with vanishing external momenta}
\label{fig6}
\end{figure}
Let us first consider the box diagram  in  Fig.\ref{fig6}
which involves two scalar propagators and two CHS propagators.
Let us recall  that the CHS propagator \rf{33}  is proportional to
the transverse-traceless (TT) projector  
$\rP^{\mu_{1}\cdots \mu_{s}}_{\nu_{1}\cdots\nu_{s}}(k)$\, satisfying 
\be\lab{510}
	k_{\mu_1}\,\rP^{\mu_{1}\cdots \mu_{s}}_{\nu_{1}\cdots\nu_{s}}(k)=0
	=\eta_{\mu_{1}\mu_{2}}\,\rP^{\mu_{1}\cdots \mu_{s}}_{\nu_{1}\cdots\nu_{s}}(k)\,.
\ee
When all external momenta vanish, the only non-vanishing momentum
in the  box diagram is  the internal momentum $k$, and $\rP_{s}(k)$
of spin $s$  CHS propagator will be necessarily contracted with $k$ 
making the diagram vanish.
Therefore, the only non-vanishing contribution to the 
local counterterm $(\phi^{*}\phi)^{2}$  may come only 
 from the diagram with \mt{s=s'=0}\,, i.e. from the contribution of the  ``non-propagating"  spin 0 
 member of the CHS tower  (with free action $\int d^4 x (h_0)^2$, cf. \rf{1111}), and is   given by 
\be\lab{511}
	A^{(1)}_{\rm\sst Box}=\left(\frac{n_{0}}{\k}\right)^{2}\,I(\L)\,.
\ee
Here $n_{s}$ is given in \rf{43} and $I(\L)$ is the  standard UV divergent loop integral,
\be\lab{512}
	I(\L)=\int^{\L} \frac{d^{4}k}{(k^{2})^{2}}\,.
\ee
%All the other types of diagrams with  spin $s\ne0$ exchanges  are finite.

\subsubsection*{CHS bubble diagram}

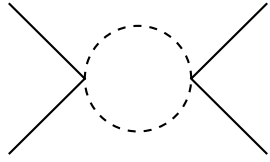
\begin{figure}[h]
\centering
\begin{tikzpicture}
\draw [thick] (0,0) -- (1,1) -- (0,2);
\draw [thick] (3.4,0) -- (2.4,1) -- (3.4,2);
\draw [thick,dashed] (1.7,1) circle [radius=0.7];
\end{tikzpicture}
\caption{Bubble diagram in scalar QED } %with $A^\m A_\m\,\phi^*\phi$ vertices}
\label{fig7}
\end{figure}

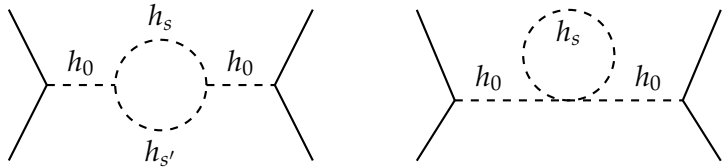
\begin{figure}[h]
\centering
\parbox{150pt}{
\begin{tikzpicture}
\draw [thick] (0,0) -- (0.5,1) -- (0,2);
\draw [thick] (4,0) -- (3.5,1) -- (4,2);
\draw [thick,dashed] (0.5,1) -- (1.4,1);
\draw [thick,dashed] (2.6,1) -- (3.5,1);
\draw [thick,dashed] (2,1) circle [radius=0.6];
\node at (0.95,1.3){$h_{0}$};
\node at (3.05,1.3){$h_{0}$};
\node at (2,1.9){$h_{s}$};
\node at (2,0.1){$h_{s'}$};
\end{tikzpicture}}
\parbox{120pt}{
\begin{tikzpicture}
\draw [thick] (0,0) -- (0.5,0.8) -- (0,2);
\draw [thick] (4,0) -- (3.5,0.8) -- (4,2);
\draw [thick,dashed] (0.5,0.8) -- (3.5,0.8);
\draw [thick,dashed] (2,1.4) circle [radius=0.6];
\node at (0.95,1.1){$h_{0}$};
\node at (3.05,1.1){$h_{0}$};
\node at (2,1.7){$h_{s}$};
\end{tikzpicture}}
\caption{Diagrams contributing to  $(\phi^{*}\phi)^{2}$ }
\label{fig8}
\end{figure}

The fact that the box diagrams with  spin  ${s\ge1}$ exchanges do not give any contribution
to UV divergence is  similar to the scalar  QED case  where 
  the UV divergence arises only  from
the bubble diagram (Fig.\ref{fig7}) with two $A^\m A_\m\,\phi^*\phi$
vertices.
In the present case of the scalar coupled to  CHS theory, we do not have 
higher order contact scalar interactions $\mathcal{O}(h^{2},\phi^{2})$ in the action \rf{1}. 
 Hence, one might  think that no one-loop bubble diagrams 
 can induce $(\phi^{*}\phi)^{2}$  term in the effective action 
  because none of them are 1-PI.
  However,  the usual distinction between 1-PI and non-1-PI diagrams 
  does not  formally apply  in $d=4$  CHS theory  due to the presence of non-propagating $s=0$   field  which has the 
 %This point is, in fact,  erroneous in the CHS theory due to
% the  presence  of non-propagating  $h_{0}$ field 
  % in the spectrum. 
 % that has simply
   $(h_0)^2$  kinetic term  (see \rf{1111}). 
 It turns out that the diagrams in Fig.\ref{fig8}
 (where the $h_0$ lines there  are  effectively shrunk  to a point and $h_s$ loops
  include also the contributions of the corresponding ghosts)
  do produce  zero-momentum  $(\phi^{*}\phi)^{2}$ terms. % for given spin $s$ and $s'$\,.
%Even if each spin contribution is not zero, there is still a chance that
%the summed contribution over all spin may vanish. 
We   shall return to the   analysis of these contributions 
in  section \ref{sec: CHS loop}. % and move on to the other diagrams.

\subsubsection*{Charge renormalisation diagrams}

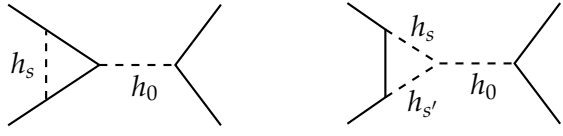
\begin{figure}[h]
\centering
\begin{tikzpicture}
\draw [thick] (0,0) -- (1.2,0.8) -- (0,1.6);
\draw [thick,dashed] (1.2,0.8) -- (2.2,0.8);
\draw [thick,dashed] (0.5,0.35) -- (0.5,1.25);
\draw [thick] (2.8,0) -- (2.2,0.8) -- (2.8,1.6);
\node at (0.2,0.8){$h_s$};
\node at (1.8,0.5){$h_0$};
\end{tikzpicture}
\qquad\qquad 
\begin{tikzpicture}
\draw [thick] (0,0) -- (0.5,0.35) -- (0.5,1.25) -- (0,1.6);
\draw [thick,dashed] (0.5,0.35) -- (1.2,0.8);
\draw [thick,dashed] (0.5,1.25) -- (1.2,0.8) -- (2.2,0.8);
\draw [thick] (2.8,0) -- (2.2,0.8) -- (2.8,1.6);
\node at (1,0.27){$h_{s'}$};
\node at (1,1.27){$h_{s}$};
\node at (1.8,0.5){$h_0$};
\end{tikzpicture}
\caption{Charge renormalization diagrams}
\label{fig5'}
\end{figure}

The ``charge renormalization" diagrams involving the one-loop correction to  the $h_s\,\phi^*\,\phi$ vertices
 may  also contribute
to the $(\phi^*\,\phi)^2$ contact term through  the $h_0$ internal line
(see  Fig.\ref{fig5'}).

%and involve  $h_s$-propagator of arbitrary spin.
%However, for the same reason that
%only $h_0$ coupling is relevant in the box diagram,
As in the case of the box diagram, here again the only non-trivial diagrams 
with constant external scalars 
are the ones which involve only $s,s'=0$ internal lines.
Moreover, it follows 
 from  dimensional analysis that the  there is  no ${h_0}^3$ vertex in the CHS  action
 so that   the only non-vanishing contribution 
  comes from  the first   diagram   in Fig.\ref{fig5'}  with $s=0$. 
Its contribution is  given by 
\be\lab{5.4}
A^{(1)}_{\rm\sst charge\, ren.}=\left(\frac{n_{0}}{\k}\right)^{2}\,I(\L)\,.
\ee

\subsubsection*{Scalar bubble diagram}

\begin{figure}[h]
\centering
\begin{tikzpicture}
\draw [thick] (0,0) -- (0.5,1) -- (0,2);
\draw [thick] (4,0) -- (3.5,1) -- (4,2);
\draw [thick,dashed] (0.5,1) -- (1.4,1);
\draw [thick,dashed] (2.6,1) -- (3.5,1);
\draw [thick] (2,1) circle [radius=0.6];
\node at (0.95,1.3){$h_{0}$};
\node at (3.05,1.3){$h_{0}$};
\end{tikzpicture}
\caption{Non 1-PI diagram with scalar loop}   
\label{fig11}
\end{figure}
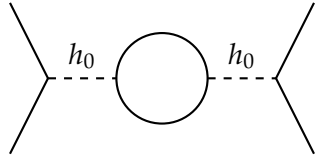

Finally, there is also  a possible contact  $(\phi^{*}\phi)^{2}$  contribution  from 
the  non 1-PI diagram   with scalar loop and non-propagating $h_0$ field  in Fig.\ref{fig11}.
%%%%%%%%%%%%%%%%%%%%%%%%%%%%%
%%%%%%%%%%%%%
We find 
\be
	A^{(1)}_{\rm\sst scalar\, bubble}=N_{\phi}\left(\frac{n_{0}}{\k}\right)^{2}\,I(\L)\,,
	\label{54}
\ee
where  for generality   we included  the factor 
$N_{\phi}$  of the number of massless scalars (in the discussion above  we   had  $N_{\phi}=1$). 
%Introduction of multiple scalars, hence the dependence of $N_{\phi}$

%\ed{/////As a preparation for computation of on-shell one-loop amplitude  for the graph in Fig.\ref{fig2},   let us  first consider 
%a  generalization of  the computation  in Section \ref{sec3} to the case when 
%two upper legs of the graph  in Fig.\ref{fig1}   are on-shell  and the two   lower  legs  are off-shell. 
%Having  found such  correlator to compute the amplitude in Fig.\ref{fig2} 
% we will need   to   combine  two of them  together  and   integrate over  the  virtual  momentum running through the loop. 
% A surprising feature of  the  amplitude  in Fig.\ref{fig2}  is   that  its imaginary part should vanish as its   cut 
% is proportional to the tree-level   4-scalar amplitude that vanishes as  shown in 
% Section \ref{sec3}. 
%[this assuming  unitarity that may not however apply here directly,  once S-matrix   contains  external CHS lines]
%Here     we  also expect to get zero result 
%for its coefficient due to summation over all spins (in an appropriate regularization)  as 
%  such term  should  be  prohibited by the  global CHS  symmetry discussed in the previous section. }

\subsection{Equivalent approach: integrating out $h_0$ first}
%%%%%%%%%%%%%%%%%%%%%%

The fact that $h_0$  is non-propagating
allows one to treat it as an auxiliary field, i.e. integrate it out 
%In the above consideration, the fact that $h_0$ 
%is auxiliary --- that is, its equation is algebraic rather than differential --- played a crucial role. 
%Since any auxiliary field can be integrated out from the classical theory
%without destroying the locality,
%we can also first integrate out  $h_0$ 
ending up with a local  action  for the remaining  fields $\phi$ and $h_{s\ge1}$\,.
 The price % of removing $h_0$ dependence,
is    getting  new interaction vertices. 
\begin{itemize}
\item
First, the CHS action itself will be modified.
Since the $h_0$ equation is of the form
$h_0=\mathcal{O}(h_{s\ge1}^2)$\,,  we get 
 additional vertices  at  quartic or higher orders.
These  will not, however,   contribute to the four-scalar scattering at the one-loop order. 
%there is no additional diagram due to these vertices.
%%%%%%%%%%%%%%%%%%%%%%%%%%%%%%%%%%%%%%%%
\item 
The presence of $\phi^{*}\phi\, h_0$ coupling in \rf{2.6} 
implies   that,  after solving for $h_0$,    the  massless 
scalar   scalar  action acquires the self interaction vertex
 $(\phi^{*}\phi)^{2}$.
As a result,   there will   be   extra diagrams in  Fig.\ref{fig14}  contributing 
to  the four scalar scattering.
These are, of course, equivalent  to 
the  $s=s'=0$ diagram  in Fig.\ref{fig6},  the first  diagram with $s=0$  in 
Fig.\ref{fig5'}  and the diagram in Fig.\ref{fig11}  with  all $h_0$ lines shrunk to a point. 
%%%%%%%%%%%%%%%%%%%%%%%%%%
\begin{figure}[h]
\centering
\parbox{100pt}{
\begin{tikzpicture}
\draw [thick] (0,2) -- (1,1.5);
\draw [thick] (1,1.5) arc (90:270:0.5);
\draw [thick] (0,0) -- (1,0.5);
\draw [thick] (2.1,2) -- (1.1,1.5);
\draw [thick] (1.1,0.5) arc (-90:90:0.5);
\draw [thick] (2.1,0) -- (1.1,0.5);
\end{tikzpicture}}
\parbox{100pt}{
\begin{tikzpicture}
\draw [thick] (0,0) -- (0.5,0.95);
\draw [thick]  (0.5,1.05) -- (0,2);
\draw [thick] (0.5,1.05) arc (175:-175:0.5);
\draw [thick] (2.1,0) -- (1.6,1) -- (2.1,2);
\end{tikzpicture}}
\parbox{100pt}{
\begin{tikzpicture}
\draw [thick] (0,0) -- (0.5,1) -- (0,2);
\draw [thick] (2.2,0) -- (1.7,1) -- (2.2,2);
\draw [thick] (1.1,1) circle [radius=0.5];
\end{tikzpicture}}
\caption{One-loop diagrams with  $(\phi^{*}\phi)^{2}$ vertices (broken or open lines 
indicate the origin of these diagrams in relation to diagrams in Figs. \ref{fig6}, \ref{fig5'}, \ref{fig11}).}
\label{fig14}
\end{figure}
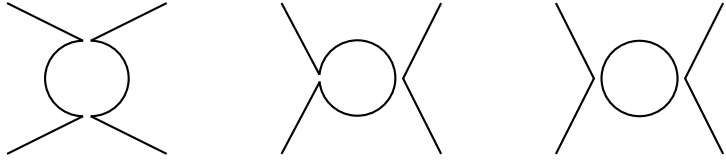
%%%%%%%%%%%%%%%%%%%%%%%%%%%
\item 
Finally, there will appear additional interaction vertices between $\phi$  and $h_{s\geq 1}$, 
notably, the vertices of type
$h^{2}\,\phi^{2}$ and
$h^{2}\,\phi^{4}$  (see  Fig.\ref{fig9}).
These  lead to  extra   one-loop diagrams in   Fig.\ref{fig10}, 
which   again  are equivalent  to 
the diagrams in Fig.\ref{fig8}  with $h_0$ lines shrunk to a point. 
%%%%%%%%%%%%%%%%%%%
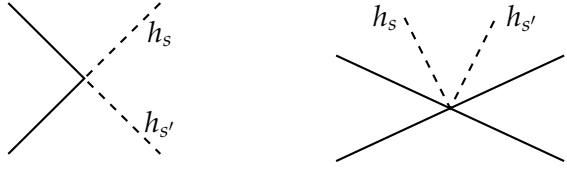
\begin{figure}[h]
\centering
\parbox{120pt}{
\begin{tikzpicture}
\draw [thick] (0,1) -- (1,0) -- (0,-1);
\draw [thick,dashed] (2,1) -- (1,0) -- (2,-1);
\node at (2,0.6){$h_{s}$};
\node at (2,-0.6){$h_{s'}$};
\end{tikzpicture}}
\parbox{80pt}{
\begin{tikzpicture}
\draw [thick] (1.5,0.7) -- (-1.5,-0.7);
\draw [thick] (-1.5,0.7) -- (1.5,-0.7);
\draw [thick,dashed] (-0.6,1.2) -- (0,0) -- (0.6,1.2);
\node at (-0.85,1.2){$h_{s}$};
\node at (0.95,1.2){$h_{s'}$};
\end{tikzpicture}}
\caption{Higher order contact vertices}
\label{fig9}
\end{figure}
%%%%%%%%%%%%%%%%%%%%%
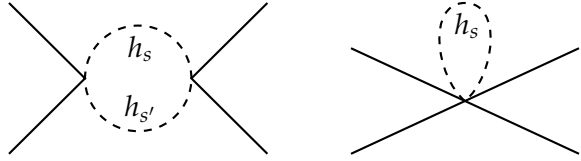
\begin{figure}[h]
\centering
\begin{tikzpicture}
\draw [thick] (0,0) -- (1,1) -- (0,2);
\draw [thick] (3.4,0) -- (2.4,1) -- (3.4,2);
\draw [thick,dashed] (1.7,1) circle [radius=0.7];
\node at (1.73,1.4){$h_{s}$};
\node at (1.73,0.6){$h_{s'}$};
\draw [thick] (7.5,1.4) -- (4.5,0);
\draw [thick] (4.5,1.4) -- (7.5,0);
\draw [thick,dashed] (6,0.7) to [out=140,in=180] (6,2) to [out=0,in=40] (6,0.7);
\node at (6.03,1.7){$h_{s}$};
\end{tikzpicture}
\caption{One-loop diagrams with
 $h^{2}\,\phi^{2}$ and  $h^{2}\,\phi^{4}$ vertices}
\label{fig10}
\end{figure}
 \end{itemize}
 %%%%%%%%%%%%%%%%%%%%%%%%%%%
 In this  approach,  with $h_0$ integrated out first, all UV divergences of the four-scalar scattering amplitudes come from two types of 1-PI diagrams: the one (Fig.\ref{fig14}) involving $\phi$-loops  and the other one  (Fig.\ref{fig10}) involving $h_s$-loops  (where in general one is  also
to add  ghost loop  contributions):
\be\lab{566}
                                A^{(1)}_{\rm tot}=A^{(1)}_{\phi\rm-loop} + A^{(1)}_{h_s\rm -loop}\,.
\ee
The former contributions were   already   given  in \eqref{511},\rf{5.4}  and \eqref{54} and  thus we find, symbolically,
\be\lab{5.6}
A^{(1)}_{\phi\rm-loop}= (1 + 1  + N_{\phi})\left(\frac{n_{0}}{\k}\right)^{2}\,I(\L)\,.
\ee
The  dependence on $N_{\phi}$  makes  it clear that $A^{(1)}_{\phi\rm-loop}$ cannot be canceled by $A^{(1)}_{h_s\rm-loop}$ 
since the latter is independent of $N_{\phi}$\,.

Thus  for  generic value of $N_{\phi}$ 
the one-loop four scalar scattering amplitude will have a UV divergence,
i.e.  the amplitude  will  not vanish contrary to  what  happened at the tree level.
This  may  not  be 
 in contradiction with the CHS global symmetry argument of Section \ref{sec4} because
scalar loop contributions  may  render the CHS symmetry anomalous.
 
 One possible approach is  to treat the  scalar $\phi$  field as an external only, i.e. 
 to ignore  all diagrams   with $\phi$  scalar loops altogether. 
 It is then  of interest to see  if  the contributions the remaining diagrams  with CHS loops only  in Fig.\ref{fig10} 
  may  vanish when  summed over all  spins. 
  This  will be addressed in the next subsection. 
  %below. 

%As explained already, this cannot be compensated by the diagrams in Fig.\ref{fig10} as the latter are
%$N_\phi$ independent.
\iffa 
Even though the one-loop amplitude does not 
vanish as opposed to the tree-level case,
we shall proceed to calculate 
the remaining diagrams in Fig.\ref{fig10} 
for the completeness. 
In fact the diagrams in Fig.\ref{fig10} has  
its own interest as we shall discuss in the next section.
\fi 

%Since $h_0$ dependence in the action is 
%$h_0^2   + h_0 \phi* \phi$, integrating out   $h_0$  we just  get local 
%$( \phi* \phi)^2$  vertex   and  then are  still to compute one-loop 
%diagram  of \ref{fig7} type with all lines  now being   scalars.
%The  result  was  already  found  in \rf{511} (up to overall coefficient). 

%%%%%%%%%%%%%%%%%%%%%%%%%%%%%%%%%%
\subsection{Divergent part of one-loop CHS  effective action  in constant $h_0$ background }
\label{sec: CHS loop}

Let us now consider   the diagrams in Fig.\ref{fig8} 
(or equivalently those in Fig.\ref{fig10})
where the  external scalar field $\phi$  lines are taken at zero  momentum
%is assumed to be constant  so is the 
(so that  same  applies to $h_0$ lines  in Fig.\ref{fig8}). 
The UV divergent contribution  from the diagrams in Fig.\ref{fig8} takes the form
\be
	c_{\rm\sst CHS}\,\left(\frac{n_{0}}{\k}\right)^{2}\,I(\L)\,\int d^4x\,(\phi^*\,\phi)^2\,,
	\label{cpp}
\ee
 where  the coefficient $c_{\rm\sst CHS}$   encodes the contributions  from infinitely many 
 CHS field loops (Fig.\ref{fig10}).
Equivalently, this 
 constant  appears in  the UV divergent  $h_0$ dependent 
 part of the one-loop effective action of the CHS theory 
\be
	\G^{(1)}_{\rm\sst div}[\,h_0]
	=c_{\rm\sst CHS}\,  I(\L)\, \int d^4x\, ({h_0})^2\,.
	\label{chh}
\ee
%%%%%%%%%%%%%%%%%%%%
\iffa 
where the full functional $\G^{(1)}_{\rm\sst div}[h]$ is defined formally by
\be 
	-\hbar\,\log\left[\int_\L Dh\,e^{-\frac1{\hbar}\,S_{\rm\sst CHS}[h+\bar h]}\right]
	=S_{\rm\sst CHS}[\bar h]+\hbar\left(\G^{(1)}_{\rm\sst fin}[\bar h]
	+I(\L)\,\G^{(1)}_{\rm\sst div}[\bar h]\right)+\mathcal{O}(\hbar^2)\,.
\ee
\fi
\def \ccc  {{\rm c}}  \def \aaa  {{\rm a}}  
%%%%%%%%%%%%%%%%%%%%%%%
On general grounds, the CHS  theory  $S_{\rm CHS}=\k \int d^4 x ( h_0^2 + F_{\m\n}^2  + C^2_{\m\n\l\r} + ....)$
having dimensionless coupling  constant  should be renormalizable (the gauge symmetries 
fix the local  action uniquely)  and thus  the same  $c_{\rm\sst CHS}\, \log \Lambda$  one-loop  coefficient should appear 
in front of the (linearised) Weyl tensor  term if  spin 2 background   is turned on in addition to $h_0$ 
in  \rf{chh}. Then $c_{\rm\sst CHS}$ should   be the  same as the  conformal  anomaly $\ccc$-coefficient
of the CHS   theory.   The conformal anomaly $\aaa$-coefficient of the CHS theory
(corresponding to topological Euler number divergence in the effective action) 
  was found in 
  \cite{Giombi:2013yva,Tseytlin:2013jya,Giombi:2014iua} to vanish   if  a natural regularization for summation over all spins  is used. The same  vanishing was  found  also    for the total 
  $\ccc$-coefficient  \cite{Tseytlin:2013jya,Beccaria:2014xda,Beccaria:2015vaa}
  under the assumption 
  that contributions  to conformal anomaly from higher derivative   CHS  operators 
  on Ricci flat background  factorize.
One may thus expect that total $c_{\rm\sst CHS}$   coefficient   of the UV divergent $h_0^2$ term in \rf{chh} should also vanish. 

%%%%%%%%%%%%%%%%%%%%%%%%%%%
\iffa 
%%%%%%%%%%%%%%
Once we have the one-loop effective action, we can study the conformal anomaly
by examining how the action transforms under the dilatation.
Since the mutual rescaling with cut-off parameter ($\delta_\s I(\L)=-\s$) should leave the action invariant:
\be
	\delta_{\s}\,\G^{(1)}_{\rm\sst fin}
	+\delta_\s I(\L)\,\G^{(1)}_{\rm\sst div}
	+I(\L)\,\delta_\s \G^{(1)}_{\rm\sst div}=0\,,
\ee
we obtain the standard properties,
\be
	\delta_{\s}\,\G^{(1)}_{\rm\sst fin}=\s\,\G^{(1)}_{\rm\sst div}\,,
	\qquad
	\delta_\s \G^{(1)}_{\rm\sst div}=0\,.
\ee
If we assume that  the conformal invariant $\G^{(1)}_{\rm\sst div}$ is also anomaly free
for the entire CHS gauge symmetry, then it should simply coincide with 
the classical CHS action itself up to an overall constant:
\be
	\G^{(1)}_{\rm\sst div}=\frac{c_{\rm\sst CHS}}{2\,\k}\,S_{\rm\sst CHS}\,,
	\label{1l c}
\ee
because it ought to be the only local functional invariant under CHS gauge symmetries
up to topological terms.
If we restrict the non-trivial background of CHS fields only to the spin-two sector, namely the metric tensor,
then $S_{\rm\sst CHS}$ will contain the Weyl tensor square term, hence corresponds to
the $c$-anomaly.
Therefore, the computation of the coefficient $c_{\rm\sst CHS}$ in \eqref{chh}
is expected to determine also the $c$-anomaly of CHS theory
if all above assumptions are correct.
The conformal anomaly of CHS theory
has been investigated in  \cite{Giombi:2013yva,Tseytlin:2013jya,Giombi:2014iua,Beccaria:2014xda,Beccaria:2015vaa}.
\fi 
%%%%%%%%%%%%%%%%%%%%%

To  check this    let us  directly 
evaluate the logarithmically divergent part of the one-loop effective action 
of CHS theory  assuming that  the only non-trivial background is the  constant 
spin 0 field  $h_0$\,.
%Since $h_0$ has the mass dimension two, the only possible structure is
%the no-derivative ${h_0}^2$ term with a coefficient $c_{\rm\sst CHS}$ as in \eqref{chh}.
%Hence, the identification of the effective action in this background boils down to the identification
%of the coefficient $c_{\rm\sst CHS}$\,. Moreover, one can further assume that the background field $h_0$ is spacetime constant. As there is no derivative term in \eqref{chh}, we do not lose any generality with this assumption.
%In the above, when we discuss the CHS bubbles, or equivalently the diagrams of Fig.\ref{fig10},
%we have omitted the ghost contribution for simplicity.
To compute  $c_{\rm\sst CHS}$ from the diagrams of Fig.\ref{fig10}
we need to take into account both the ``physical" (gauge-fixed)  field loop  and the ghost loop contributions, i.e.
%Hence, the coefficient  can be split into two pieces:
\be\lab{5.13}
	c_{\rm\sst CHS}=c^{\sst ph}_{\rm\sst CHS}+c^{\sst gh}_{\rm\sst CHS}\,.
\ee
%where $c^{\sst ph}_{\rm\sst CHS}$ receives the contribution from  
%the loops of physical fields
%while $c^{\sst gh}_{\rm\sst CHS}$ from the ghost loops.
%In the following, we shall calculate these two coefficients separately.

\subsubsection{Physical field loop contribution}
Let us first consider the loop diagrams involving physical fields.
%%%%%%%%%%%%%%%%%%%%%%%%%%%%%%
 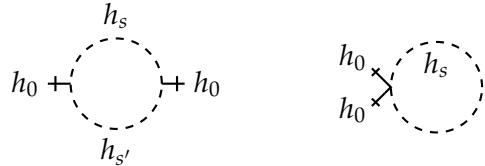
\begin{figure}[h]
\centering
\parbox{120pt}{
\begin{tikzpicture}
\draw [thick] (1.1,1) -- (1.4,1);
\draw [thick] (2.6,1) -- (2.9,1);
\draw [thick] (1.2,0.9) -- (1.2,1.1);
\draw [thick] (2.8,0.9) -- (2.8,1.1);
\draw [thick,dashed] (2,1) circle [radius=0.6];
\node at (0.8,1){$h_{0}$};
\node at (3.2,1){$h_{0}$};
\node at (2,1.9){$h_{s}$};
\node at (2,0.1){$h_{s'}$};
\end{tikzpicture}}
\parbox{100pt}{
\begin{tikzpicture}
\draw [thick] (1.15,1.65) -- (1.4,1.4) -- (1.15,1.15);
\draw [thick] (1.15,1.55) -- (1.25,1.65);
\draw [thick] (1.15,1.25) -- (1.25,1.15);
\draw [thick,dashed] (2,1.4) circle [radius=0.6];
\node at (0.9,1.8){$h_{0}$};
\node at (0.9,1.1){$h_{0}$};
\node at (2,1.7){$h_{s}$};
\end{tikzpicture}}
\caption{CHS effective action in $h_0$ background}
\label{fig15}
\end{figure}
%%%%%%%%%%%%%%%%%%%%%%%
There are two types of 1-PI diagrams in Fig.\ref{fig15}
and their evaluation requires the knowledge of $h_0\,h_s\,h_{s'}$ and 
${h_0}^2\,{h_s}^2$ vertices.
These vertices can be represented in momentum space as
\be\lab{f1}
\parbox{80pt}{\begin{tikzpicture}
\draw [thick,dashed] (-0,0) -- (1,0);
\draw [thick,dashed] (1.5,0.8) -- (1,0) -- (1.5,-0.8);
\node at (-0.4,0){$h_{0}$};
\node at (1.8,0.9){$h_{s}$};
\node at (1.8,-0.9){$h_{s}$};
\end{tikzpicture}}=\k\,
\tilde h_{0}(0)\,C_{s}(k,\partial_{u_{1}},\partial_{u_{2}})\,
\tilde h_{s}(k,u_{1})\,\tilde h_{s}(-k,u_{2})\,\big|_{u_{i}=0}\,,
\ee
\be\lab{f2}
\parbox{80pt}{\begin{tikzpicture}
\draw [thick,dashed] (-0.8,0.8) -- (0.8,-0.8);
\draw [thick,dashed] (0.8,0.8) -- (-0.8,-0.8);
\node at (-1.1,0.9){$h_{0}$};
\node at (-1,-0.9){$h_{0}$};
\node at (1.1,0.9){$h_{s}$};
\node at (1.1,-0.9){$h_{s}$};
\end{tikzpicture}}=\k\,
\tilde (h_{0}(0))^{2}\,Q_{s}(k,\partial_{u_{1}},\partial_{u_{2}})\,
\tilde h_{s}(k,u_{1})\,\tilde h_{s}(-k,u_{2})\,\big|_{u_{i}=0}\,.
\ee
Here the two functions $C_{s}$ and $Q_{s}$
 encode all tensor structures:
\be
	C_{s}(k,\partial_{u_{1}},\partial_{u_{2}})= c_{s}\,(k^{2})^{s-1}
	\left(\partial_{u_{1}}\!\cdot\partial_{u_{2}}\right)^{s}
	+...\ , \quad 	Q_{s}(k,\partial_{u_{1}},\partial_{u_{2}})= q_{s}\,(k^{2})^{s-2}
	\left(\partial_{u_{1}}\!\cdot\partial_{u_{2}}\right)^{s}
	+...\ .
	\label{vCQ}
\ee
Here  dots  stand for  terms involving at least one trace or one divergence of a field 
so that they  drop out in 
the traceless and transverse gauge that we shall assume.
% such terms  drop hence 
%the only relevant information is on the sequence $c_s$ and $q_s$\,. 
For the same reason  we can consider only $h_0\,(h_s)^2$ vertices
instead of more general $h_0\,h_s\,h_{s'}$ ones because the latter 
necessarily contain a trace or divergence.

Using the vertices \eqref{vCQ} we get,  respectively, for 
 the  left and the  right diagram in Fig.\ref{fig15} 
\ba\lab{f3}\no 
	&&I_1=\frac{1}{4}
	\left(\frac{n_{s}}{\k}\right)^{2}
	\int d^{4}k\left.\frac{
	\k\,C_{s}(k,\partial_{u_{1}},\partial_{u_{2}})\,
	\k\,C_{s}(k,\partial_{v_{1}},\partial_{v_{2}})\,
	\rP_{s}(k,u_{1},v_{1})\,\rP_s(k,u_{2},v_{2})}
	{(k^{2})^{2s}}\right|_{u_{i}=v_{i}=0},\\
%\ee
%and
%\be\lab{f4}
	&&I_2=\frac{1}{4}\,
	\frac{n_{s}}{\k}
	\int d^{4}k\left.\frac{
	\k\,Q_{s}(k,\partial_{u_{1}},\partial_{u_{2}})\,
	\rP_{s}(k,u_{1},u_{2})}
	{(k^{2})^{s}}\right|_{u_{i}=0}\,,
\ea
where we have used the propagator \eqref{33} involving the traceless and transverse projector $\rP_s$. 
After removing the  auxiliary variables $u_i$ and $v_i$ (which amounts to
the contraction of all the indices) and performing the $k$-integral 
(which reduces to  the UV divergent term  \rf{512}),
we obtain
\be\lab{5.18} 
	I_1=\te \frac{1}{4}(2s+1)\, 
	\left(n_{s}\,c_{s}\right)^{2}\,I(\L)\,,
	\qquad\qquad 
	I_2=\frac{1}{4}\, (2s +1) \,
	n_{s}\,q_{s}\, I(\L)\,.
\ee
Here the factor $2s+1$ comes from the trace of the projector $\rP_s$
(this is the dimension of the symmetric rank $s$ representation of $so(3)$
which is the traceless and transverse part of 4d Lorentz tensor).
The cubic \eqref{f1} and quartic interactions \eqref{f2},
or equivalently  the coefficients $n_{s}\,c_{s}$ and $n_{s}\,q_{s}$
in \rf{5.18} can be extracted from the CHS action. 
This  is done in   Appendix \ref{app2} and  with the  result being 
\be\lab{5.19}
\te 	n_{s}\,c_{s}=-4\left(s+\frac12\right)\ ,  \quad s\ge1\,  ;
	\qquad\qquad \qquad 
\te 	n_{s}\,q_{s}=8\left(s+\frac12\right)\left(s-\frac12\right)\ ,\quad s\ge2\,.
\ee
Finally, using  the above  expressions,  we  obtain
\be\lab{5.20}
	c^{\sst ph}_{\sst\rm CHS}=2^{3}\sum_{s=1}^{\infty}{\te \left(s+\frac12\right)^{3}
	}  +2^{2}\sum_{s=2}^{\infty}\te \left(s+\frac12\right)^2\left(s-\frac12\right) \ .
\ee
The sum over spins is formally divergent and thus requires an appropriate 
definition or regularization to be discussed below.

%Hence, in order to get a meaningful result, we need again a suitable
%regularization.
%We shall come back to this issue after 
%identifying the ghost contribution.

%%%%%%%%%%%%%%%%%%%%%%%%%%%%%%%%%%%%%%
\subsubsection{Ghost loop contribution}
%%%%%%%%%%%%%%%%%%%

To find the  ghost  contribution  corresponding to  the traceless transverse gauge 
let us  consider the  gauge   symmetries of the  classical CHS action. 
%The CHS theory in four dimensions extends the theory of Maxwell and Weyl gravity to higher spins,
%hence admits gigantic amount of gauge symmetries.
%In the tree-level and the physical field loop calculations, we have used the traceless and transverse gauge without further discussions.
%A proper treatment of CHS gauge fixing will introduce ghost fields which contribute to the one-loop effective action of CHS theory. 
%, for example via the diagram showed on figure \ref{fig12}.
%\begin{figure}[h]
%\centering
%\parbox{150pt}{
%\begin{tikzpicture}
%\draw [thick] (0,0) -- (0.5,1) -- (0,2);
%\draw [thick] (4,0) -- (3.5,1) -- (4,2);
%\draw [thick,dashed] (0.5,1) -- (1.4,1);
%\draw [thick,dashed] (2.6,1) -- (3.5,1);
%\draw [thick,dotted] (2,1) circle [radius=0.6];
%\node at (0.95,1.3){$h_{0}$};
%\node at (3.05,1.3){$h_{0}$};
%\node at (2,1.9){$c_{s}$};
%\node at (2,0.1){$c_{s'}$};
%\end{tikzpicture}}
%
%\caption{1-Loop ghost diagram contributing to $(\phi^{*}\phi)^{2}$ amplitude}
%\label{fig12}
%\end{figure}
Since we are interested in  computing the one-loop 
 ghost contribution in  a constant $h_0$ background,  it is sufficient  to consider 
%Effectively, this reduces the 
the classical CHS action to quadratic order in all $s >0$ fields, i.e.
%into a quadratic one 
with $h_0$-dependent kinetic operator
\be\lab{5.21}
	S_{\rm\sst CHS}
	=\int d^4x\,\la h|K(h_0)|h\ra\,,
\ee
where $\la\cdot|\cdot\ra$ stands for the contraction of indices.
When the background $h_0$ is turned off,
the operator $K$ reduces to that of the free CHS theory.
The above action is invariant under
the  following gauge transformation (cf. Appendix \ref{appsym})
\be	\delta_{\e,\a}\, h=u\cdot\partial_x\, \epsilon + 
	\left[u^2
	- h_0\,\cF(\partial_u,\partial_x)\right] \alpha\,,
	\label{h0 tr}
\ee
where the gauge fields and parameters 
can be chosen to be doubly-traceless and traceless, respectively, without  loss of generality.
The $h_0$ dependent part
of gauge transformation is given with
the operator $\cF(\partial_u,\partial_x)=
\Pi_d(\partial_u,\partial_x)\,\Pi^{-1}_{d+4}(\partial_u,\partial_x)$\,.
In the following, we shall gauge fix the CHS  field $h$ to
traceless and transverse one  by making use of 
the transformation \eqref{h0 tr}.

First, using the $\alpha$ part of the transformation \eqref{h0 tr}, 
we can   gauge fix the trace of $h$
to zero.
This step does not introduce any ghost (since the transformation is algebraic)
but modifies the residual gauge transformation to the form
\be 
	\delta_{\e}\,h=T(h_0,\e)=
	\rP_{\rm\sst T}\left[u\cdot\partial_x
	- h_0\,\cG(\partial_u,\partial_x)\right] \epsilon\,,
	\label{R g}
\ee
where $\rP_{\rm\sst T}$ is the traceless projector
and the precise form of $\cG(\partial_u,\partial_x)$ is
given in Appendix \ref{appgaugefix}.
Due to the tracelessness of the parameter $\epsilon$\, 
the gauge transformation \eqref{R g} remains linear in 
$h_0$ even after this traceless gauge fixing.

Second, using  the remaining transformation \eqref{R g},
we can  make the traceless field $h$   also transverse.
 This step involves differential part of the gauge transformation
 which gives  rise to a non-trivial Jacobian.
 The latter can be  represented 
 by an appropriate  ghost contribution 
 (see  Appendix \ref{appgaugefix}), with 
 the ghost action being 
 \be
 	S_{gh}=\int d^4x\,\la \bar c | K_{gh}(h_0)| c\ra\,,
 	\label{ghost}
 \ee
% with the operator,
 \be\lab{521}
 	K_{gh}(h_0)=\partial_x\cdot\partial_u\frac{\delta T(h_0,\e)}{\delta \e}=\partial_x\cdot\partial_u\,
 	\rP_T\,
	\left[u\cdot\partial
	- h_0\,\cG(\partial_u,\partial_x)\right].
\ee
 The crucial observation is that %all the degrees of freedom encoded in the $h_0$ dependent terms are also encoded in the $h_0$ independent  term. This means that
  by  shifting appropriately the ghost  fields $c$  one  can   completely eliminate all $h_0$ dependence. 
  This is to be done spin by spin, starting with 
   the lower spin,  so that each ghost field is shifted once  and then left alone. 
   % The bottom line is that  in this way the $h_0$ 
% dependence can be completely eliminated from the ghost action \eqref{ghost}.
It is important to note that the existence of this redefinition is due
to an additional divergence term in 
 the operator of gauge transformation.
The gauge transformation itself cannot be
redefined in such a way that it becomes independent of $h_0$\,.
The details of this  argument  are given in Appendix \ref{appgaugefix}.

We thus  conclude that 
the CHS ghosts do not couple to a constant $h_0$ background, and 
hence %its one loop cannot contribute to the ${h_0}^2$ term,
\be\lab{522}
	c^{\sst gh}_{\rm\sst CHS}=0\,.
\ee
%This is rather unexpected because there is no a prior reason that the ghost contribution
%should vanish.

%%%%%%%%%%%%%%%%%%%%%%%%%
\subsubsection{Summing over spins}  %and Regularization}
%%%%%%%%%%%%%%%%
The  final  expression   for the coefficient of the divergent $h_0^2$ term 
in the CHS   action may thus be written as  (see  \rf{chh},\rf{5.13},\rf{5.20},\rf{522})
\be\lab{523}
	c_{\rm\sst CHS}= - 5 
	%\frac12+\frac92
	+4 \sum_{s=0}^{\infty}\te \left[ 3\left(s+\frac12\right)^{3}
	-\left(s+\frac12\right)^2\right] \ .
\ee
%as the proportionality constant of the divergent part in the one loop CHS effective action \eqref{1l c}.
As was  mentioned above,  this coefficient   should be expected to be   proportional
to the  conformal  anomaly $\ccc$-coefficient of the CHS theory.
The   expression  for the  latter   was  found to vanish
 \cite{Tseytlin:2013jya,Giombi:2014iua,Beccaria:2014xda}
 provided the sum over spins   is defined   using the $e^{-(s+ \a)\,\e}$ cutoff  with $\a= \ha$
just as in the   similar vanishing of the 
$\aaa$-coefficient  \cite{Giombi:2013yva,Tseytlin:2013jya}.\foot{The  computation 
of the one-loop  conformal anomaly $\ccc$-coefficient in the CHS theory  is based on  two assumptions: 
(i)   the CHS action obtained as an induced action  in near-flat space expansion can be 
reformulated (using  a field redefinition)  in such a way that  at least    quadratic kinetic terms 
 in generic  curved metric background  are     reparametrization and Weyl invariant;
 (ii) the higher derivative kinetic   operators $D^{2s} + ...$  -- while not factorizing, in general, 
 into products of $D^2+...$ operators in a Ricci-flat  background \ci{Nutma:2014pua}
 (as they do in   $AdS$ or sphere  background) --
 still  contribute to $\ccc$-anomaly  in the same  way as  if they do factorize 
 (the  terms with derivatives of the curvature tensor that obstruct factorization cannot contribute to $C^2_{\m\n\k\l}$ 
 conformal anomaly on dimensional grounds).}
The same  cutoff   factor  $e^{- (s + \alpha_d) \e}$  appeared in 
 \rf{45}   with $\alpha_d= {d-3 \ov 2} = \ha$ in $d=4$. 
Using    such  exponential cutoff  $e^{-(s+ \a)\,\e}$ in \rf{523}   and dropping all 
singular terms we get 
%Several  regularisations can be examined butthey give  different results in general.
%For instance, if one introducesthe regulator $e^{-(s+a)\,\e}$ (the zeta function regularization corresponds to the choice $a=1/2$),then we obtain 
\be 
	c_{\rm\sst CHS}(\a) =\te 
\frac{1}{30} \left(90\, \a^4-140\, \a^3+75\, \a^2-15\, \a- 152\right),
\ee
which does not, however,  vanish for a rational value of $\a$\,.

The  meaning of this   observation is unclear  at the moment. 
One possibility is that \rf{523} is missing   some contributions  making it different from the
 conformal anomaly $\ccc$-coefficient  discussed in \cite{Tseytlin:2013jya,Beccaria:2014xda,Beccaria:2015vaa}.
Indeed, the CHS   spin $s$ field  conformal anomaly coefficients  are 6-th order polynomials\foot{Explicitly, $
\aaa_s = { 1 \ov 720} \nu_s ( 3 \nu_s + 14 \nu_s^2)$ \cite{Giombi:2013yva,Tseytlin:2013jya}
and  $
\ccc_s - \aaa_s = { 1 \ov 720} \nu_s ( 4- 45 \nu_s + 15 \nu_s^2)$   where  $\nu_s = s(s+1)$.}
  while the summand in \rf{523} is only cubic polynomial in $s$.
  At the same time, the partial  spin $s$ contributions to $h_0^2$  and $C^2_{\m\n\k\l}$ 
  divergent terms in the CHS action need not match:   what   is expected to  be the same 
  is only the total summed over spins  coefficients.

Another possibility is   that while $c_{\rm\sst CHS}$ in \rf{523} is not related to the conformal
anomaly $\ccc$-coefficient  its still  determines the UV divergent  part of the CHS loop   contribution  to the four-scalar   amplitude. In that  case to resolve the  regularization ambiguity 
it would   be important (as in the tree-level amplitude case in \rf{314}) to keep the external momentum non-zero (which would play, e.g.,  the role of $z$ in \rf{48}).\foot{In the case of  the  one-loop partition function 
of the massless  higher spin theory in AdS 
 it was noticed \ci{Giombi:2014iua}
that  a  consistent   result can be obtained
by first summing over spins and  then removing the UV cut-off. 
In our case as well, first summing over the spins and  then sending  momentum $p$ to zero may
lead  to  
more  sensible result
than  first   setting momentum to zero and then summing over spins.}
Equivalently, 
%Hence, in order to have a better control on our result, it is 
it would be desirable to 
repeat the calculation of  the CHS effective action in a \emph{non-constant} $h_0$ background.
%considering the spin summation at the last stage.

%%%%%%%%%%%%%%%%%%%%%%%%%%%%%%%%%%%%

%It would  amount to identifying the terms which depend on the external momentum $p$
%of $\tilde{h}_0(p)$  ignored in the current section \ref{sec: CHS loop}.
%Related to the latter point, let us make one more remark. 

%%%%%%%%%%%%%%%%%%%%%%%%%%%%%%%%%%%%%%%%%%%%%%%%
\section{Concluding remarks}	
\label{Concl}
%%%%%%%%%%%%%%%%%%%%%%%

The $d=4$ conformal higher spin theory  having vanishing  total  coefficients of
 the  conformal anomaly  (and thus possibly of all higher symmetry anomalies) 
 is  a potentially  consistent   quantum theory of  an infinite tower of  higher spin fields   having a large amount of symmetry. While apparently non-unitary due to higher derivatives in the
 $s>1$ kinetic terms  this theory has a well-defined   formulation  in flat space  background   and thus 
deserves  a    detailed  investigation. 
 
Here we have studied the   scattering  amplitudes   for a massless  conformal scalar $\p$ coupled to CHS theory. The four-scalar   tree-level amplitude is given by   the exchange of  the whole tower  of CHS  fields. 
We have found   that under a natural prescription of summation  over spins 
the resulting tree level amplitude   vanishes  
%v3
for generic physical momenta. 
This vanishing turns out to be 
  in agreement   with  the expectation  based on global 
extended   conformal symmetry. 

We also addressed the  extension of this computation to one-loop order. 
We considered only the simplest   case of  vanishing external  scalar momenta.
The one-loop   diagrams  contributing  to the four-scalar scattering  are of two types: 
(i)  involving  internal scalar  propagators (i.e.  scalar loops), and (ii) involving only CHS  field  loops. 
The former are potentially anomalous (scalar loop in external  CHS background  has, in particular, a 
 non-vanishing Weyl anomaly)  and thus   the symmetry  argument  of section 3  about the vanishing 
 of the  total amplitude due to   global CHS symmetry need not apply.  
 We have thus   concentrated on  the   CHS loop contributions only. The expectation is 
   that the  coefficient $c_{\rm \sst CHS}$ of the UV divergent term 
    in the zero-momentum amplitude (or  of the $(\phi^*\phi)^2$  term in the effective action) 
 should   be the same   as the  conformal anomaly $\ccc$-coefficient 
 and  should thus vanish   after summation over all spins. %    assuming appropriate summation 
  The  expression  for $c_{\rm \sst CHS}$   we have found  does not vanish however
 and that issue 
requires further investigation. 
It would be important, in particular, to clarify the  precise relation 
between the coefficient $c_{\rm\sst CHS}$   found in section 5 and the conformal
 anomaly $\ccc$-coefficient.\foot{To recall, 
 the main  logical steps were as follows. 
 The  coefficient $c_{\rm \sst CHS}$   we computed  was the   coefficient of the 
 $h_0^2 \log \Lambda$ term in the CHS effective action. The conformal  anomaly $\ccc$-coefficient 
 is the same as the  coefficient of the $C^2_{\m\n\l\r}   \log \Lambda$ term in the CHS effective action. 
 The full  UV  divergent   term in the one-loop effective action   should  be  invariant not only 
 under the Weyl symmetry and reparametrizations  but also under the whole CHS gauge symmetry.
 As  there is a  unique local functional which is invariant under CHS gauge symmetry 
 \ci{Segal:2002gd,Bekaert:2010ky}, the  coefficients of the $h_0^2 $ and $C^2_{\m\n\l\r}$ terms 
 are thus  expected to be the same.}

  %and to verify the existence of a regularization of the sum over spins 
%which  implies the vanishing of the CHS one-loop  correction to the external 
 %four-scalar amplitude.
 
 It would  be interesting also to   apply the methods of the present paper 
 to the computation of the  tree and one-loop  S-matrix    for the CHS   fields  themselves
 (e.g., Maxwell vector and Weyl graviton). That may  provide  further   evidence for the existence  of a 
  consistent regularization of the sum over spins  and may also       shed   some  light  on the 
 (non)unitarity  issue.

 \iffa 
 In principle, the logic is not something proven. We assume
1)      c anomaly is given by the log divergent part in 1-loop.

2)      The anomaly term is not only invariant under the Weyl+diffeo but also under the whole CHS gauge symmetry.

3)      There is unique local functional which is invariant under CHS gauge sym.

The point 1 is a matter of definition hence should be fine. The point 3 is more or less proven by Segal (and with Jihad and Xavier, we also got the same conclusion). The point 2 is not examined much, but it seems correct to my intuition.
 
I checked a few times the calculation of physical loop part and I don�t find any error (I had made two errors in the beginning. The factor of 2 in the propagator and (2s+1) factor from the TT projector).

 \fi

%$\aaa$-  and possibly  $\ccc$-

%From physical points of view, the expected non-unitarity is the main obstacle that needed to overcome
%in order CHS theory to be considered as a physically relevant theory. 
%Despite of the apparent perturbative non-unitarity, we may still study various physical processes
%to see which kind of problem CHS theory has (or is free from).
%For such a reason, in this paper, we calculate the amplitudes of several scattering processes
%involving CHS interactions.
%More precisely, we consider the scattering of conformal scalars via the exchange of CHS particles. 

\iffa 
%%%%%%%%%%%%%%%%%%%%%
We consider  set-up analogous to    4d AdS/CFT with   probe brane degrees of freedom  kept and scattered   while  large $N$ 
 number of     CFT multiplets integrated out inducing  massless  HS    theory in the bulk or CHS  theory at the boundary. 
 What we  are going to consider is analog of  boundary  gluon S-matrix at strong coupling. The difference  with SYM is that   there  S-matrix is nontrivial due to interactions in SYM theory while here scalar theory is free.  Then it is   in a sense   not surprising that we should end up with trivial S-matrix 
 %%%%%%%%%%%%%%%%%%%%%%
\fi

%%%%%%%%%%%%%%%%%%%%%%%%%%%%%%%%%%%%%%%%%%
\acknowledgments
We  thank  M. Grigoriev,   R. Metsaev, J. Mourad, D. Ponomarev, R. Roiban  and M. Taronna 
 for useful discussions. 
The work of EJ   and AAT  was  supported 
 by  the  Russian Science Foundation grant 14-42-00047  associated with Lebedev Institute.
 The work of EJ was supported also by Basic Science Research Program through the National Research Foundation of Korea (NRF) funded by the Ministry of Education(2014R1A6A3A04056670).
The work of  SN   and AAT was   supported  by  the STFC ST/L00044X/1.
The work of AAT was   also  supported   by   the 
 % STFC Consolidated  grant ST/J0003533/1   and the 
ERC Advanced grant No.290456.

\def \ha {{\te {1\ov 2}}}
\appendix

\iffa 
%%%%%%%%%%%%%%%%%%%%%%%%%%%%%%%%%%%%%%%%%%%
\section{Another  example of regularization of the sum over spins}
%%%%%%%%%%%%%%

In section 3.2 we considered two prescriptions of how to define the sum of the 
CHS exchange   contributions over all spins.   
Here   we shall describe  yet another  regularization, which  is based on  the use of  zeta-function.
Let   us  first expand the Legendre polynomials in \rf{48}  in Taylor series near $z=-1$ as
\ba\lab{52}
	P_{s}(z)=\sum_{n=0}^{\infty} \frac{(z+1)^{n}}{n!}\,P^{[n]}_{s}(-1)\,,
\ea
and then sum each monomial over $s$\,. Using the expression for the $n$-th  derivative of  the Legendre polynomial
we get 
\be\lab{53}
	P^{[n]}_{s}(-1)=(-1)^{s}\,\frac{2^{n}}{n!}\,\prod_{m=1}^{n}{\te \Big[\big(s+\frac12\big)^{2}-
	\big(m-\frac12\big)^{2}\Big]}
	=(-1)^{s}\sum_{m=0}^{n}\,p_{m,n}\,\te \big(s+\ha\big)^{2m}.
\ee
Then we can represent  $F_{4}$ in \rf{48}  as
\be
	F_{4}(z)=\sum_{n=0}^{\infty} f_n \frac{(1+z)^{n}}{n!}\,
	\label{F4}
\ , \qquad \qquad 
	f_{n}=\sum_{m=0}^{n}\,p_{m,n}\,\sum_{s=0}^{\infty}(-1)^{s}\big(s+\ha\big)^{2m+1}.
\ee
%Notice that in this way, the function $F_{4}(z)$ is transformed to the sequence $f_{n}$\,.
Each $f_{n}$ involves divergent sum over $s$  which needs to be regularized.
It turns out to be crucial that the above series involves  only odd powers of $(s+\ha)$\,.
The $f_{n}$  can be regularized as 
\be\lab{57}
	f^{\rm reg}_{n}=\sum_{m=0}^{n} p_{m,n}\,\Phi\big(-1,-2m-1,\ha\big) \ , 
\ee
 where $\Phi$   is   the Lerch zeta-function 
\be\lab{56}
	\Phi(x,y,a)=\sum_{s=0}^{\infty}\frac{x^{s}}{(s+a)^{y}}\,.
\ee
Applying  the relation   between the  Lerch and  Hurwitz zeta function $\zeta (y,a)$\,,
\be\lab{58}
	\Phi(-1,y,a)=2^{-y}\left[\zeta\big(y,\frac{a}2\big)-\zeta\big(y,\frac{a+1}2\big)\right],
\ee
one can show that 
\be\lab{59}
	\Phi\big(-1,-2m-1,\ha\big)
	=\frac{2^{2m+1}}{2(m+1)}
	\left[ B_{2(m+1)}\big(\frac34\big)-B_{2(m+1)}\big(\frac14\big)\right]=0\,,
\ee
where $B_{n}(x)$ is Bernoulli polynomial  having the property $B_{n}(1-x)=(-1)^{n}\,B_{n}(x)$\,. 
Hence, we get the vanishing $f^{\rm reg}_{n}$   and thus the  vanishing  of $F^{\rm reg}_{4}(z)$\,, 
i.e. the same result  as found in section 3.2  using  other 
 regularisations.

\fi

%%%%%%%%%%%%%%%%%%%%%%%
%%%%%%%%%%%%%%%%%%%%%
\section{Review  of  global CHS symmetry } \label{appsym}

Let us  review  the origin of the  global CHS symmetry  and its action on the  free scalar and CHS fields 
 following \ci{Segal:2002gd,Bekaert:2010ky}.

%\ed{[[ Check all factors in this section ]]}
\subsection*{Transformation of massless  scalar field}

The   massless scalar action \rf{2.1}  may be written in  the following  \emph{operator} representation 
\be
	S_{\rm\sst free}[\phi]=\la \phi|\,\hat p^{2}\,|\phi\ra\,,
	\label{433}
\ee
where \mt{\phi(x)=\la x|\phi\ra} and \mt{\hat p_{\mu}=i\,\partial_{\mu}}\,.
To find  the maximal symmetries of this action  we consider 
the most general transformation linear in $\phi$\,. In the operator formulation, it reads
\be
	\delta\,|\phi\ra=i\,\hat t\,|\phi\ra\,,
	\label{444}
\ee
where $\hat t$ is  an arbitrary  polynomial in $\hat x$ and $\hat p$, i.e. a  differential operator  acting on $\phi(x)$\,.
The condition that it preserves the action \eqref{433} is
\be
		\qquad \hat p^{2}\,\hat t=\hat t^{\dagger}\,\hat p^{2}\,.
		\label{preserve}
\ee
This defines the maximal symmetries of conformal scalar action up to the \emph{trivial} ones
\be
\lab{446}	\delta \phi^{i}=C^{ij} (\phi) \,\frac{\delta S_{\rm\sst free}}{\delta \phi^{j}} \ , \qquad \qquad  C^{ij}=-C^{ji}  \ , 
\ee
 which  are proportional to the equations of motions, i.e. 
   vanish  on-shell.
Such trivial transformations correspond  in  the  case of \rf{433}  to the operator  of the form
\be
	\hat t=\hat r\,\hat p^{2}\ ,  \qquad \qquad  \hat r^{\dagger}=\hat r\,,
	\label{quotient}
\ee
with  $\hat r$ an arbitrary hermitian factor.
The set of operators $\hat t$ satisfying \eqref{preserve} modulo \eqref{quotient}
defines the global CHS symmetry  that  acts on conformal scalars as in \eqref{444}.\foot{
In fact, the global CHS symmetry in $d$ dimensions is nothing but the Vasiliev's HS algebra in $(d+1)$ dimensions.
The typical formulation of Vasiliev's HS algebra involves differential operators in $(d+2)$-dimensions, while here we formulated it 
in terms of differential operators in $d$-dimensions. The reason for the  existence of  the two descriptions is
the fact that the conformal scalar in $d$-dimension can be formulated in $(d+2)$-dimensions where  the role of $\hat p^{2}$ is played by
the  three operators
$
	\hat X^{2},\ \  2(\hat X\cdot\hat P+\hat P\cdot \hat X),\ \  \hat P^{2}, 
$
which form an  $\mathfrak{sp}(2,\mathbb R)$ algebra.
See \ci{Joung:2014qya} for a recent overview of the HS algebra. }

A convenient way to treat the operators is by using the Wigner-Weyl correspondence (see, e.g., 
  appendix A in \ci{Bekaert:2010ky}). Then we can map the operator $\hat t$
 to a phase-space function \be \lab{477}
 t(x,p)=e(x,p)+i\,a(x,p)\equiv (e, a)\,,  \ee
  and all operator products become Moyal products.
  In this formulation the conformal scalar transforms as
\be
	\lab{488}
	\delta\phi(x)=e^ {- \frac i2 \partial_{x_{2}}\cdot\partial_{u}}
	t\, (x_{1},u)\,\phi(x_{2})\,\Big|_{_{\overset{x_{1}=x_{2}=x}{\sst u=0}}}\,, 
\ee
  where 
the conditions  on $\hat t$  in \rf{477}  to represent the    CHS symmetry  are 
 \ba
 	&p\cdot\partial_{x}\,e-(p^{2}+\partial_{x}^{2})a=0\,,	\label{CHS sym}\\
	&
	(e,a) \sim (e,a)+ \left((p^{2}+\partial_{x}^{2})\,r\,,\,p\cdot\partial_{x}\,r\right).\label{equiv}
\ea 
The algebraic structure is  induced from the operator product as (here  the commutators in the r.h.s. are defined  using the  Moyal ${}^\star$ product)
\be\lab{4111}
	\Big[\,\big(e_{1}\,,\,a_{1}\big)\,,
	(e_{2}\,,\,a_{2})\,\Big]
	=\big(\ [\, e_{1}\,\overset{\star},\,e_{2}\,]
	- [\, a_{1}\,\overset{\star},\,a_{2}\,]\,,\,
	[\, e_{1}\,\overset{\star},\,a_{2}\,]
	+ [\, a_{1}\,\overset{\star},\,e_{2}\,]\ \big)\,.
\ee
The global CHS symmetry contains the conformal algebra  with generators
\be
	P_{\mu}=(p_{\mu},0)\,,\qquad M_{\mu\nu}=(x_{[\mu}\,p_{\nu]},0)\,,
	\qquad
	K_{\mu}=(x_{\mu}\,x\cdot p, x_{\mu})\,,
	\qquad
	D=(x\cdot p,1)\,,
\ee
and also other higher spin  generators, for example, the generators of  \emph{hyper-translations}
\be
	P_{\mu_1 ...\mu_r} = 	(p_{\{\mu_{1}} ... p_{\mu_{r}\}},\ 0)\,, \lab{4133}
\ee
where $\{...\}$   indicates the subtraction of all traces.

\subsubsection*{Transformation of   CHS fields}

The   above   symmetry   may be also considered as  
a
global  part of  the gauge   symmetry    acting   on  the CHS  fields.
It  will then be a symmetry of the action of the  free scalars coupled to the CHS fields  as  in \rf{2.6}.

The action of this conformal higher spin  symmetry on the CHS fields becomes  more transparent  in the 
so-called dressed formulation \ci{Segal:2002gd}, where  one uses a different set of CHS fields $\mathfrak{h}(x,u)$
which  related to the original one  \rf{299}  
 by \be\lab{4144}
   \mathfrak{h}(x,u)=\Pi_{d} (\partial_u, \del_x) \,h(x,u) \ , \ee
    where $\Pi_{d}$   was defined in  \rf{25}
 (see \ci{Bekaert:2010ky} for details). 
%\be 	\mathfrak{h}^{\sst (s)}(x,u)=  \sum_{n=0}^{\infty}\,   \frac{1}{n!\,(s+\frac{d-3}{2}+n)_{n}}  \left(\ell^{2\,}\frac{(\partial_{u}\cdot\partial_x)^2
 %   -\partial_{u}^2\,\partial_x^2}{16}\right)^{n}\,     h^{\sst (s+2n)}(x,u)\,,\ee 
The    CHS action \rf{301} then  becomes  a non-diagonal functional
\be\lab{416}
	S_{\rm\sst CHS}[\mathfrak{h}]=
	\int d^{d}x\ U_{\frac{d-3}2}\Big(  (\partial_{x_{12}}\cdot\partial_{u_{12}})^{2}-\partial_{x_{12}}^{2}\,\partial_{u_{12}}^{2} \Big)\,\mathfrak{h}(x_{1},u_{1})\,\mathfrak{h}(x_{2},u_{2}) \Big|_{_{\overset{x_{1}=x_{2}=x}{\sst u_1=u_2=0}}} \ 
	+\mathcal{O}(\mathfrak{h}^{3})\ ,
\ee
where $U_{\nu}(z)=(\sqrt{z}/2)^{-\nu}\,J_{\nu}(\sqrt{z}/2)$\, \ ($J_\nu$ is a Bessel function). The advantage of working with $\mathfrak{h}$ is that the CHS gauge symmetry takes  a simple form
\be
\lab{418}	\delta\,\mathfrak{h}=[\, \mathfrak{e}\ \overset{\star},\ u^{2}+\mathfrak{h}\,]
	+\{\,\mathfrak{a}\ \overset{\star},\ u^{2}+\mathfrak{h}\,\}
	=\delta^{\sst (0)}\mathfrak{h}+\delta^{\sst (1)}\mathfrak{h}\,, 
\ee
where $\star$ acts on the space of functions in $x$ and $u$\,.
$\delta^{\sst (0)}$ and $\delta^{\sst (1)}$ are respectively   $\mathfrak{h}$-independent
and $\mathfrak{h}$-linear parts, and 
the gauge parameters are related to those in  \eqref{2999} by
\ba
    \mathfrak{e}(x,u) \eq  \Pi_{d+2}(\partial_{u},\partial_{x})\ \epsilon(x,u)
    +(\partial_x\cdot\partial_u)\ \Pi_{d+2}(\partial_{u},\partial_{x})\
    \frac{1}{2(d-1)+4\,u\cdot\partial_{u}}\,\alpha(x,u)\,,\nn
    \mathfrak{a}(x,u)\eq\Pi_{d+4}(\partial_{u},\partial_{x})\ \alpha(x,u)\,.
    \label{4188}
\ea
The field-independent part of the transformation reads
\be
\lab{419}	\delta^{\sst (0)} \mathfrak{h}
	\te =u\cdot\partial_{x}\,\mathfrak{e}+
	\left(u^{2}-\frac14\,\partial_{x}^{2}\right) \mathfrak{a}\,.
\ee
This coincides with the  l.h.s. of \eqref{CHS sym} and
the equivalence relation \eqref{equiv} can be interpreted here as  a  ''{gauge for gauge}'' symmetry,
\be
\lab{420}	\te \delta\mathfrak{e}=\left(u^{2}-\frac14\,\partial_{x}^{2}\right)r\,,
	\qquad
	\delta\mathfrak{a}=-u\cdot\partial_{x}\,r\,.
\ee
Hence for the special parameter $(\mathfrak e,\mathfrak a)=(e,a)$
 satisfying $\delta^{\sst (0)}\mathfrak{h}=0$\,
(which can be interpreted as  the conformal Killing equation \eqref{CHS sym})
 the CHS action \rf{416}
is invariant under 
\be
\lab{421}	\delta\, \mathfrak{h}=
	[\,e\ \overset{\star},\ \mathfrak{h}\,]
	+\{\,a\ \overset{\star},\ \mathfrak{h}\,\}\,.
\ee
This defines the action of the global CHS symmetry on the CHS fields. Since it acts linearly, 
it preserves all different $\mathfrak{h}^{n}$-parts of the  CHS action separately; 
in particular, it leaves  its  quadratic  part in \rf{416}  invariant. 

The interaction  \rf{2.6} between the CHS fields $\mathfrak{h}$ and the conformal scalar  
  (with  currents  written in  the \emph{un-dressed} form, cf. \rf{31},\rf{999})  
\be\lab{422} 
	S_{\rm\sst int}[  \p, h]=\int d^{d}x\,\mathfrak{h}(x,\partial_{u})\,\mathfrak{J}(x,u)\,\big|_{u=0}\ ,
\ee
is also invariant under the global CHS symmetry.  
This  becomes manifest by writing it in the  operator form as 
\be\lab{4.24} 
	S_{\rm\sst int}[\phi,h]=\la \phi|\,\hat{\mathfrak{h}}\,|\phi\ra,
\ee
where $\hat{\mathfrak{h}}$ is the operator corresponding to the symbol $\mathfrak{h}(x,p)$\,.

\def \half {{1\ov 2}}

%%%%%%%%%%%%%%%%%%%%%%%%%%%%%%
\section{Cubic and quartic vertices in the CHS action  involving  constant $h_{0}$ field}
\label{app2}

Let us
start with   %remind the reader that
recalling that given  the  heat kernel expansion for the massless scalar  kinetic operator 
in conformal higher spin background,
\be\lab{c1}
	{\rm Tr}\left[ e^{-t\left(\hat p^{2}+\hat{\mathfrak{h}}\right)}\right]
	=  \sum_{n=0}^{\infty}t^{n-2}\,a_{n}[\mathfrak{h}]\,, 
\ee
 the local  CHS action  in $d=4$   can be defined as the second Seeley coefficient 
 (i.e. as  the  coefficient of the logarithmic UV divergence in the induced action) 
\be
	S_{\rm\sst CHS}[\mathfrak{h}]\propto a_{2}[\mathfrak{h}]\,.
\ee
Let us separate  the spin-0 part of CHS field $\mathfrak{h}_{0}$
from the rest  of the fields $\mathfrak{h'}$\,:
\be
	\mathfrak{h}(x,u)=\mathfrak{h}_{0}(x)+\mathfrak{h'}(x,u)\,.
\ee
Here $\mathfrak{h}(x,u)$ is defined in \rf{299},\rf{4144} 
(the distinction between $\mathfrak{h}(x,u)$  and ${h}(x,u)$
will not be important in traceless transverse gauge). 
Then restricting $\mathfrak h_{0}$ to be constant one obtains
\be
	a_{n}[\mathfrak{h}]=\sum_{m=0}^{\infty}\,\frac{(-1)^{m}}{m!}\,
	(\mathfrak h_{0})^{m}\,a_{n-m}[\mathfrak{h'}]\,.
\ee
In particular,
\be\lab{c4}
	a_{2}[\mathfrak{h}]=a_{2}[\mathfrak{h}']
	-\mathfrak h_{0}\,a_{1}[\mathfrak{h}']
	+\ha\,(\mathfrak h_{0})^{2}\,a_{0}[\mathfrak{h}']
	+\mathcal{O}(\mathfrak h_{0}^{3})\,.
\ee
The heat kernel coefficients $a_{n}$ were calculated in \cite{Bekaert:2010ky}
up to quadratic order in $\mathfrak{h}$\,,
\ba
	a_{2+m}[\mathfrak h]
	\eq\int \frac{d^{4}x}{(4\pi)^{2}}\,
	{\te \sqrt{\frac{\pi}8}\,
	\left(\frac12 {\partial_{x_{12}}^{2}}\right)^{\!m}}\, 
	U_{m+\half}\left((\partial_{x_{12}}\!\cdot\partial_{u_{12}})^{2}
	-\partial_{x_{12}}^{2}\partial_{u_{12}}^{2}\right)\nn
	&&\hspace{100pt}\times\,\mathfrak{h}(x_{1},u_{1})\,\mathfrak{h}(x_{2},u_{2})\,\Big|_{_{\overset{x_{1}=x_{2}=x}{\sst u_1=u_2=0}}}
	%\Big|_{u_{i}=0}
	+\mathcal{O}(\mathfrak{h}^{3})\,,
\ea
\ba
	a_{1-m}[\mathfrak h]
	\eq\int \frac{d^{4}x}{(4\pi)^{2}}\Big[\te 
	\delta_{m,1}+\left(\frac14\,\partial_{u}^{2}\right)^{m}\mathfrak h(x,u)\Big|_{u=0}\nn
	&&\hspace{50pt}+\, \te 
	\sqrt{\frac{\pi}8}\,V_{m}(\partial_{x_{12}},\partial_{u_{12}})\,
	\mathfrak{h}(x_{1},u_{1})\,\mathfrak{h}(x_{2},u_{2})\,\Big|_{_{\overset{x_{1}=x_{2}=x}{\sst u_1=u_2=0}}}
	%\Big|_{u_{i}=0}
+\mathcal{O}(\mathfrak{h}^{3})\Big]\,,
\ea
where 
\be
	V_{m}(\partial_{x},\partial_{u})
	={\te \left(\frac14\,\partial_{u}^{2}\right)^{m+1}}
	\sum_{k=0}^{\infty}\te \frac{\left(\frac18\,\partial_{x}^{2}\,\partial_{u}^{2}\right)^{k}}
	{\Gamma(k+m+2)}\,U_{k+\half}\left((\partial_{x}\cdot\partial_{u})^{2}\right) \ , 
\ee
and $U_{\nu}(z)$  is the same as in \rf{416}, i.e.  %is defined as follows:
\be
	U_{\nu}(z)={\te \left(\frac{\sqrt{z}}2\right)^{-\nu}\,J_{\nu}\left(\frac{\sqrt{z}}2\right)}
	=\sum_{m=0}^{\infty} \te \frac1{m!\,\Gamma(\nu+m+1)\,2^{\nu}}\left(-\frac{z}{16}\right)^{m}\,.
\ee
As a result, the CHS Lagrangian  depending on 
  constant $\mathfrak{h}_{0}$
and traceless and transverse $\mathfrak{h}'$   and written in   momentum space reads
($\mathfrak{h} (x) \to\tilde{ \mathfrak{h}}(p)$)
\ba
	\td { \cL}_{\sst\rm CHS}[\mathfrak{h}]
	&\!\propto\!& 
	\sum_{s=0}^{\infty}
	\Big[1-
	{4\ov p^2}\left(s+\ha\right){\tilde{\mathfrak{h}}_{0}(0)}
	+{8\ov p^4}\left(s+\ha\right)\left(s-\ha\right)\, \big({\tilde{\mathfrak{h}}_{0}(0)}
\big)^2   	+\mathcal{O}\left(\tilde{\mathfrak{h}}_{0}^{3}\right)\Big]\nn
&&\qquad\times\,
	\frac{\left(p^{2}\right)^{s}\,\tilde{\mathfrak{h}}_{s}(p,\partial_{u})\,
	\tilde{\mathfrak{h}}_{s}(-p,u)}{\,2^{3s}\, \Gamma(s+\frac32)}
	+\mathcal{O}\left(\tilde{\mathfrak{h}}'{}^{3}\right)\,,
\ea
where $\tilde{\mathfrak{h}}_{s}(p,u)= {1\ov s!} \tilde{\mathfrak{h}}_{\mu_1...\mu_s}(p) u^{\mu_1}...u^{\mu_s}$. 
Here  the non-local terms with negative powers of $p^{2}$ should be discarded.
Hence the cubic  $\mathfrak{h}_{0}\, \mathfrak{h}_{s}^{2} $ terms start from $s=1$
where as the quartic $\mathfrak{h}_{0}^{2}\,\mathfrak{h}_{s}^{2}$ terms  start from $s=2$\,.

%%%%%%%%%%%%%%%%%%%%%%%%%%

\iffa 
I did a quick calculation. In (C.5), we need to insert P_t in front of the summation.
In (C.6) and (C.7), the factor (s+2-r)/2(s+3) should be modified to
-(s+2-r)[s^2+(3-r)s+3r+1]/[2(s+2)(s+3)]
It will be nice if Simon can check this quickly.
For the calculation I used the projector
P_t=1-u^2 \partial_u^2/(4s+ u^2 \partial_u^2)

\fi 

\section{Gauge fixing  and ghost action } \label{appgaugefix}

In this Appendix we shall discuss the  ghost action corresponding to the traceless transverse gauge on CHS fields.\foot{A discussion of an alternative gauge leading to simple
gauge-fixed action for free conformal higher spin  fields in flat space  and  the corresponding  ghost fields
may be found in ref.\ci{Metsaev:2014vda}.}
 As we have shown in Appendix \ref{appsym}, the CHS gauge symmetry takes a more concise form \eqref{418} in ``dressed" basis of fields  (defined   by \rf{4144},\rf{25}, \rf{299}).
It  is thus  more convenient  to fix the gauge in that basis. After all, in the transverse traceless (TT) 
gauge we will use, the two bases become equivalent\,: $\mh(x,u)|_{_{\rm TT}} = h(x,u)|_{_{\rm TT}}$.
In addition, the scalar parts coincide with each other, $\mathfrak{h}_0=h_0$, independently of the gauge choice.

Restricting to the  case where the only non-trivial background is constant $h_0$\,,  the  symmetry  transformation \eqref{418} reduces to the form,
\begin{equation}
	\delta\,\mathfrak{h}(x,u) = u\cdot \partial_{x}\,\mathfrak{e}(x,u) + 
\te 	\left(u^2 - \frac{1}{4}\,\partial_x^2+ h_0\right)\mathfrak{a}(x,u)\,,
	\label{hgs}
\end{equation}
where the fields $\mathfrak{h}$ are doubly-traceless while the parameters $\mathfrak{e}$ and $\mathfrak{a}$ are traceless.
We  first  gauge fix $\mathfrak{h}$ to be traceless utilizing the algebraic part of the symmetry \eqref{hgs} generated by $\mathfrak{a}$\,. 
Let us note that this gauge fixing requires in principle a finite transformation rather than an infinitesimal one. In fact, the transformation \eqref{hgs} is symmetry of the classical action  \eqref{5.21} even for finite  parameters due to its quadratic nature. Imposing $\partial_u^2(\mathfrak h+\delta_\mathfrak\,\mathfrak{h})=0$\,, we get the relation between $\mathfrak{a}$ and $\mathfrak{e}$ as
\begin{equation}
	\mathfrak{a}(x,u) = -\frac{1}{2(2+u \cdot \partial_u)} \,\partial_x \cdot \partial_u\, \me(x,u)\,,
\end{equation}
and the traceless CHS fields transform now as 
$\d\,\mh=T(h_0,\me)$ with
\begin{equation}
T(\mathfrak{h}_0,\me)(x,u) = \rP_{\rm\sst T}\Big(u \cdot \partial_x + \frac{\frac{1}{4}\,\partial_x^2 - h_0}{2(2+u \cdot \partial_u)}\,\partial_{u} \cdot \partial_x\Big)\me(x,u)\,. \label{symwithT2}
\end{equation}
Here, $\rP_{\rm\sst T}$
 is the traceless projector  which is 
$\rP_{\rm\sst T} = 1      -    { u^2(\del_u)^2   \over    4(s-2)+  2d +  u^2( \del_u)^2}$
when acting on a spin $s$ tensor.

Next, let us  further gauge fix the traceless CHS field to make it also transverse  by using the transformation \eqref{symwithT2}. Following the standard  Faddeev-Popov procedure, this step of transverse gauge-fixing introduces the ghost action
\ba
S_{gh} \eq 
\int d^4 x\,\big\la \bar c\,\big|\,\partial_x\cdot\partial_u\,\frac{\delta T(h_0,\me)}{\delta\,\me}\,
	\big|\,c\big\ra\nn
\eq \int d^4 x\,
\sum_{s=0}^{\infty} \big\la \bar c_s\big|\, 
\partial_x\cdot\partial_u\,\rP_{\rm\sst T}
\Big(u \cdot \partial_x \big|c_s\big\ra+ 
	\frac{\frac{1}{4}\,\partial_x^2 - h_0}{2(s+3)}\,\partial_{u} \cdot \partial_x\,
	\big|c_{s+2}\big\ra\Big) \ . 
 \label{ghaction1}
\ea
Here $c(x,u)=\sum_{s=0}^{\infty}c_s(x,u)$ with $c_s(x,u)=\frac1{s!}\,c_{\mu_1\cdots \mu_s}(x)\,u^{\mu_1}\cdots u^{\mu_s}$ is  the generating function for  the ghost fields
and $\la a|b\ra=\frac1{s!}\,a_{\mu_1\cdots \mu_s}\,b^{\m_1\cdots \m_s}$ is the index contraction.
Since the gauge parameter $\mathfrak{e}$ is traceless, the ghost $c$ and antighost $\bar c$ are both traceless.

For further analysis, we decompose the ghost $c$ into traceless transverse (TT) components as
\be
	c_s(x,u)=\rP_{\rm\sst T} \sum_{r=0}^s (u\cdot\partial_x)^{s-r}\,c_{s,r}(x,u)\ , \qquad
	\qquad \qquad \,\partial_u^2\,c_{s,r}=0=\partial_x\cdot\partial_u\,c_{s,r}\,.
\ee
By plugging this decomposition for $c_s$ and $c_{s+2}$ into the action \eqref{ghaction1},
one can observe that
the first two TT components
$c_{s+2,s+2}$ and $c_{s+2,s+1}$ of $c_{s+2}$ drop out in the summand. We thus  end up with
\ba\lab{c6}
	S_{gh}\eq
 \int d^4 x\,
\sum_{s=0}^{\infty}\sum_{r=0}^s\,\big\la \bar c_s\big|\, 
\partial_x\cdot\partial_u\,\rP_{\rm\sst T}\,(u \cdot \partial_x)^{s+1-r}
%\nn
%&&\hspace{60pt}\times\,
\Big( \big|c_{s,r}\big\ra+ 
	k_{s,r} \big(\te \frac{1}{4}\,\partial_x^2 - h_0\big)\partial_x^2\,
	\big|c_{s+2,r}\big\ra\Big)\ , \ \ \ \ \ 
\ea
where 
\be \lab{c66}
k_{s,r}={ (s-r+2)(s+r-3)\ov  4(s+2)(s+3) } \ . \ee
As follows from  \rf{c6}, 
one can thus  completely remove the $h_0$ dependence in the ghost action by  the ghost field redefinition
\be
	c'_{s,r}=c_{s,r}+ \te  k_{s,r} \big(\frac{1}{4}\,\partial_x^2 - h_0\big)\partial_x^2\,
	c_{s+2,r}\,.
\ee
For a fixed $r$\,, this redefinition acts as a matrix 
which changes the value of $s$\,.
Since the form of this matrix is an upper triangular one
with the identity diagonal elements,
the corresponding Jacobian is simply one.
The conclusion is that  the ghost  determinant  contribution is trivial, i.e. does not depend on $h_0$.

\bibliographystyle{JHEP}

\bibliography{CHS-Biblio}

\end{document}

/Users/arkadytseytlin/Dropbox/CHS scattering2015/CHS-Biblio copy.bib